\pdfoutput=1
\documentclass[aps,groupedaddress,superscriptaddress,twocolumn]{revtex4-1}
\usepackage{graphicx}
\usepackage{dcolumn}
\usepackage{bm}
\usepackage{amssymb,amsfonts,amsmath}

\begin{document}

\title{How to infer relative fitness from a sample of genomic sequences}
\author{Adel Dayarian}
\affiliation{Kavli Institute for Theoretical Physics,University of California, Santa Barbara, CA}
\author{Boris I.~Shraiman}
\affiliation{Kavli Institute for Theoretical Physics,University of California, Santa Barbara, CA}
\affiliation{Department of Physics, University of California, Santa Barbara, CA}

\begin{abstract}
Mounting evidence suggests that natural populations can harbor extensive fitness diversity with numerous genomic loci under  selection. It is also known that genealogical trees for populations under selection are quantifiably different from those expected under neutral evolution and described statistically by Kingman's coalescent. While differences in the statistical structure of genealogies have long been used as a test for the presence of selection, the full extent of the information that they contain has not been exploited. Here we shall demonstrate that the shape of the reconstructed genealogical tree for a moderately large number of random genomic samples taken from a fitness diverse, but otherwise unstructured asexual population can be used to predict the relative fitness of individuals within the sample. To achieve this we define a heuristic algorithm, which we test {\it in silico} using simulations of a Wright-Fisher model for a realistic range of mutation rates and selection strength. Our inferred fitness ranking is based on a linear discriminator which identifies rapidly coalescing lineages in the reconstructed tree. Inferred fitness ranking correlates strongly with  actual fitness, with a genome in the top 10\% ranked being in the top 20\% fittest with false discovery rate of 0.1-0.3 depending on the mutation/selection parameters. The ranking also enables to predict the genotypes that future populations inherit from the present one. While the inference accuracy increases monotonically with sample size,  samples of 200 nearly saturate the performance. We propose that our approach can be used for inferring relative fitness of genomes obtained in single-cell sequencing of tumors and in monitoring viral outbreaks.
\end{abstract}

\maketitle

\section{Introduction}

Most of mutations are believed to have minimal effects on the fitness of the organism and much of the analysis of the genomic data on populations (see \cite{excoffier2006computer} for a review of methods) has been based on the neutral hypothesis,  according to which the dynamics of genetic polymorphisms and the overall genetic diversity of the population are governed by the neutral {\it drift}, i.e. stochastic fluctuations in the frequency of mutations. The neutral model assumes that deleterious mutations are eliminated  by selection fast enough to not significantly contribute to population diversity and  beneficial mutations are rare enough to produce only occasional adaptive {\it sweeps}, where the population is taken over by the offspring of the adaptive genotype, transiently suppressing neutral genetic diversity. Statistical properties of genealogies generated by neutral dynamics in asexual populations are understood in great detail \cite{hein2005gene,wakeley2008coalescent} in terms of the Kingman's {\it coalescent} process \cite{kingman1982coalescent} which follows the ancestors of the present population back in time as far as the {\it Most Recent Common Ancestor} (MRCA). The neutral coalescent \cite{hein2005gene,wakeley2008coalescent} forms the basis for estimating mutation and recombination rates and provides the null hypothesis in tests for the presence of selection \cite{tajima1989statistical,fu1993statistical}.
 
Yet, as advances in sequencing have made it possible to obtain quantitative data on genetic diversity, numerous studies have reached the conclusion that non-neutral polymorphisms are ubiquitous in populations across the spectrum of life:
from viruses \cite{miralles1999clonal,moya2004population,neumann2010evolution,neher2010recombination}  and bacteria \cite{barrick2009genome} to flies \cite{sella2009pervasive}, from mitochondria \cite{seger2010gene} to cells in cancerous tumors \cite{merlo2006cancer}.  In addition, laboratory evolution experiments in bacteria \cite{lenski1991long} and yeast \cite{kao2008molecular,lang2011genetic} 
have demonstrated directly that large asexual populations contain numerous sub-clones that are continuously generated by mutation and compete for fixation. Thus, large asexual population cannot be assumed selectively neutral. 

The presence of selection affects the shape of genealogical trees, often giving them a ``comb-like'' appearance that is strikingly different from the neutral trees described by the Kingman's coalescent \cite{hein2005gene,wakeley2008coalescent,seger2010gene}. An example of the ``genealogical  anomalie'' - i.e. large deviations from neutral genealogical structure \cite{maia2004effect} - is provided by the recent study \cite{seger2010gene} of mitochondrial diversity in three distinct populations of whale lice, {\it Cyamus ovalis}, where the authors demonstrate that the observed genealogies are statistically consistent with a non-neutral model with frequent mutations of small selective effect.

\begin{figure}[t]
\includegraphics{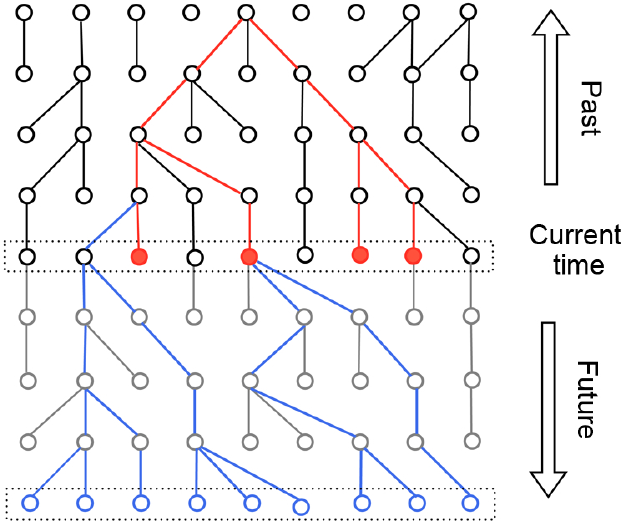}  
\caption{Schematic example of a genealogical trajectory, from past into the future, of an asexual population with fixed size ($N=9)$ and non-overlapping generations.  Nodes represent individual genomes, each linked to its ancestor in the previous generation. The example illustrates coalescence of the lineages of the bottom population towards its MRCA within the top population. The genealogical tree of a random sample (red)  from the  "current time" population partially overlaps the genealogy of the future population (blue). While actual ancestors of the future population (shown in blue) may or may not fall into the current sample, one can still define sample members that are closest to the surviving lineages. Identifying close relatives of future populations is the goal of our study.
 \label{wright_fisher}} \end{figure} 
 
Our analysis will be based on a similar model of asexual evolutionary dynamics driven by small deleterious and beneficial mutations. In Fig. \ref{wright_fisher} we show schematically a sample of continuous genealogy of a fixed size population governed by the Wright-Fisher dynamics \cite{hein2005gene,wakeley2008coalescent} incorporating genetic drift, mutation and natural selection. The example in Fig. \ref{wright_fisher} covers the period over which the offspring of one of the genomes in the top population take over the whole population at the bottom of the figure. We ask, given a sample of genomes from the ``present time'' population (shown in Fig. \ref{wright_fisher} as red discs), can one predict genetic future of the population? Or more specifically, can one identify, within the present sample, the closest relatives of the future population: i.e. individuals that are on, or closest to, the genealogical backbone of the future population? Since long term survival is correlated with fitness, this task is closely related to the problem of identifying the fitter fraction of the present day sample. 

Below, we shall demonstrate that the anomalous structure of the genealogical tree reconstructed for the sample of genomes can serve not only as the evidence for the action of selection, but also as the basis of inference of the relative fitness of sampled individuals  and their sequence closeness to the fittest genomes. Information pertinent to this inference is contained in the pattern of coalescence experienced by different lineages. In nutshell: lineages which undergo a lot of coalescence much before others are relatively fit, while the less fit lineages do not merge with the rest of lineages until they have undergone a series of mutations (backward in time). 

Our study builds on considerable recent progress in the theoretical understanding of natural selection and drift dynamics in fitness-diverse asexual populations \cite{tsimring1996rna,rouzine2003solitary,desai2007beneficial,rouzine2008traveling,o2010continuous,walczak2012structure,sniegowski2010beneficial,goyal2011rare,good2012distribution}  and the emerging description of corresponding genealogies \cite{bolthausen1998ruelle,brunet2007effect,berestycki2009recent,seger2010gene,o2010continuous,walczak2012structure,desai2012genetic,neher2012genealogies}. We shall focus on the asexual case and address how the approach might be extended to the analysis of recombining populations in the discussion.

We are interested in the regime where several beneficial or deleterious mutations segregate simultaneously and the population is formed by several clones with diverse fitness values. In this regime, often referred to as clonal interference, the interference between mutations plays an important role in determining the fate of mutations. In particular, the fate of a new mutation depends not only on its own selective effect, but also on the fitness of the genotype on which it occurs \cite{good2012distribution}. In addition, the MRCA in a fitness diverse population is with high probability among the very fittest of its generation \cite{o2010continuous}. In return, the pattern of genealogical coalescence is controlled by the time it takes for surviving lineages to converge, as they are tracked back in time, on the leading edge of the fitness distribution at previous times. 

In order to understand the parameter regime in which the interference between mutations becomes significant, consider the following  (see \cite{desai2007beneficial} for further details and proof). Let $\mu_b$ denote the beneficial mutation rate, $N$ the population size and $s$ the fitness effect of beneficial mutations. When the population size or mutation rate is small enough, the time it takes for  a new mutation to reach Þxation is less than the time it takes for another new mutation to occur and reach significant size. If a new beneficial mutation reaches the size of order $1/s$ individuals, it will escape the drift with high probability and from that point grows as $\frac{1}{s}\exp(st)$ with time, $t$. The mutation reaches the size of order $1/s$ individuals in roughly $1/s$ generations and becomes fixated in the order of $\frac{1}{s}\log(Ns)$ generations. Therefore, the total time for a mutation to reach fixation from the initial time of its occurrence is in the order of $\frac{1}{s}+\frac{1}{s}\log(Ns)$ generations. The probability for a mutation to escape drift is roughly $s$. Since the mutations are generated at rate $N\mu_b$, the time it takes for a mutation destined to escape drift to be generated is  $1/(N\mu_b s)$. 

If the parameters are such that $N\mu_b s  <<  s/(1+\log(Ns))$, a new mutation that occurs and sweeps will do so long before the next mutation destined to sweep is occured. On the other hand, for higher mutation rates or population sizes where the above inequality is not satisfied, new beneficial mutations arise and reach significant size before earlier ones can sweep, causing them to interfere with one another. For the case of purifying selection where deleterious mutations are present, it can be shown \cite{walczak2012structure} that the required condition is $N\exp(-\mu_d/s)<<\frac{1}{s}\log(\mu_d/s)$, where $\mu_d$ is the deleterious mutation rate and $s$ is the deleterious effect of mutations. For smaller $\mu_d$ or higher $s$, the deleterious mutations are purged out of the population fast enough such that the effect of selection can be captured by a simple effective population-size approximation.

\begin{figure*}[t]
\includegraphics{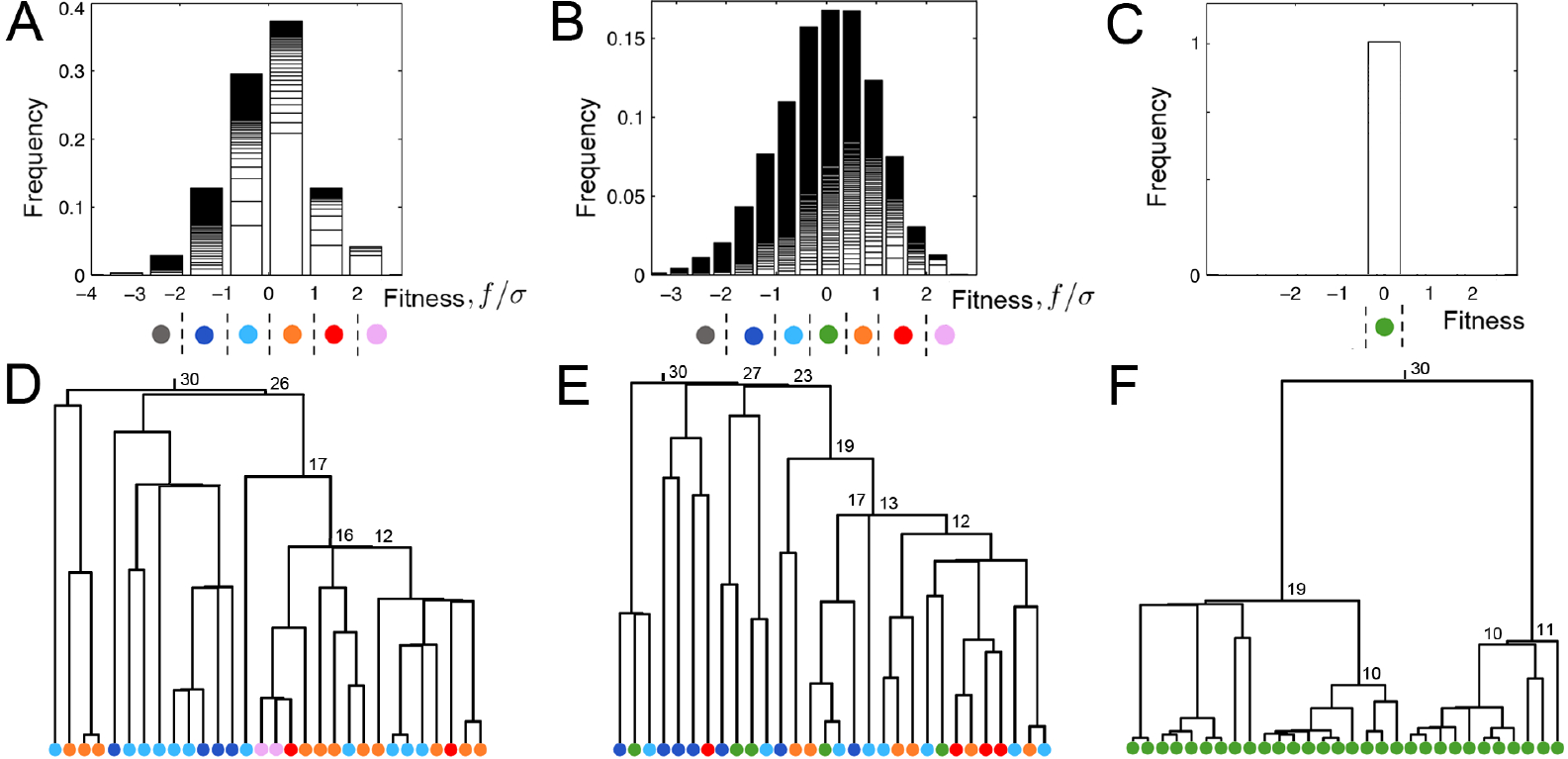}  
 \caption{
Fitness distributions and examples of genealogical trees. (A) Fitness distribution at one time point for a population with $\mu=10^{-3}$, $s=2*10^{-3}$. Each bin corresponds to a fitness class and each class is composed of multiple clones (delineated by horizontal lines within each bar, with larger clones stacked on the bottom). Clones are defined using only the non-neutral mutations. Also shown at the bottom is the color-code used in (D). (B) Same as (A) but for a higher mutation rate $\mu=10^{-2}$. The color-code used in (E) is shown at the bottom. (C) Same as (A) but for a neutral population.  (D) A typical genealogical tree for a random sample of size $n=30$ from the same population as (A). Each circle corresponds to one sampled genome and the color represents its fitness.  Branch lengths are drawn in linear proportion to the corresponding time interval. Numbers next to internal nodes are the weights of the corresponding ancestors (only weights $>10$ are shown). Note the striking asymmetry of branching, as the weight decreases in small steps along the lineage with weights marked. (E) Same as (D) but for the population shown in (B). Note that the colors (grey and plum) corresponding  to the extremes of the distribution (B) are absent from the small sample shown. (F) Same as (D) but for a neutral population. Note the short terminal legs and more symmetric branching. $N=64000$ and $\epsilon=0.1$ for all the panels. 
 \label{fit_dis}} \end{figure*}

This paper is organized as follows. After formulating the model, we shall i) provide examples of genealogies illustrating their anomalous shape as compared to the neutral coalescent and ii) demonstrate the correlation between the ancestral {\it weight}, defined as the fraction of  the present day sample constituted by the descendants of the ancestor, and the mean fitness of the those descendants. We shall then define a fitness-ranking score based on the suitably integrated ancestral weights along the reconstructed lineage of each individual in the sample. Applying the ranking to numerous sampling realizations (for populations with the same and with different mutation/selection parameters) and comparing each realization to the true fitness known from the forward simulation, we demonstrate the ability of the proposed algorithm to infer the relative fitness of sampled genomes and to identify genotypes that are likely to survive into the future. The Discussion will address possible applications and generalizations of the proposed inference method.

\section{Model}
We consider an asexual population of size $N$ that evolves with non-overlapping generations under the influx of deleterious and beneficial mutations. New mutations arise at the rate $\mu+\mu_0$ (per genome per generation) with a fraction $\epsilon \mu$ being beneficial,  $(1-\epsilon ) \mu$ deleterious and the remainder $\mu_0 $ being neutral.  For simplicity we assume both beneficial and deleterious mutations to have the same effect size $s \ll 1$ changing the fitness additively: $F_i \rightarrow F_i \pm s$.  As in the Wright-Fisher model, natural selection acts by biasing the probability of an individual genome to appear in the next generation, which is taken to be proportional to $\exp (f_i)$ with $f_i = F_i - \bar{F} $ being the individual fitness relative to  the mean fitness of the population $\bar{F}$, which in general is a function of time. 

We carried out $10^{3}$ simulations of $2*10^5$ generations for $N=64,000$ and several plausible parameter combinations in the range of $\mu =10^{-4} -10^{-2}$ and $s=10^{-3}-10^{-2}$, with  $\epsilon = 0.1$ and $0$ and $\mu_0=10\mu$.  The genealogical trees were constructed in two ways. We recorded the genealogies in the course of the forward simulation, providing exact ancestries of any sample in the population. In addition, an inferred genealogy of random samples  (between 30-500 genomes)
was constructed using standard neighbor joining/UPGMA-derived methods detailed in the Supplementary Materials (SM). 

\section{Results}
In the parameter range considered, simulated population exhibit substantial fitness diversity with fitness variance in the order of $\sigma = \big(\frac{1}{N}\sum\limits_{j=1}^{N} f_{j} ^2\big)^{1/2} \approx 10^{-3}-10^{-2}$  arising from about $10-10^{3}$ simultaneously segregating non-neutral polymorphisms.  Figure \ref{fit_dis}A-B shows examples of the population-wide fitness distribution for two different mutation rates (see SM for additional examples). In general, genetic diversity in the population is an increasing function of $\mu/s$. For the highest mutation rate and lowest selection coefficients considered, $\mu=10^{-2}$ and $s=10^{-3}$, the population is in a weak selection regime corresponding to high genetic diversity,   Fig. \ref{fit_dis}B. Lower mutation rates, as in Fig. \ref{fit_dis}A, exhibit a more clonal structure: evolutionary dynamics in this regime can be thought of as competition between multiple mutant clones.
 
Figure \ref{fit_dis}D and E show typical examples of genealogical trees constructed for random samples of size $n=30$ drawn from the populations corresponding to Fig. \ref{fit_dis}A and B, respectively. The fitness of sampled genomes, which we know from the forward simulation, is visualized using color. Also shown are ancestral weights along some of the lineages. This weight, $w_i$, is defined as the number of genomes in the sample which are direct descendants  of genome $i$. For example, each leaf at the bottom carries weight \textit{w=1}, while the node at the top carries the full weight of the sample $n=30$. For the sake of comparison, we have also shown a typical genealogical tree for a neutrally evolving population in Fig. \ref{fit_dis}F.

\subsection{Distortion in the shape of trees in the presence of selection}
\begin{figure}
\includegraphics{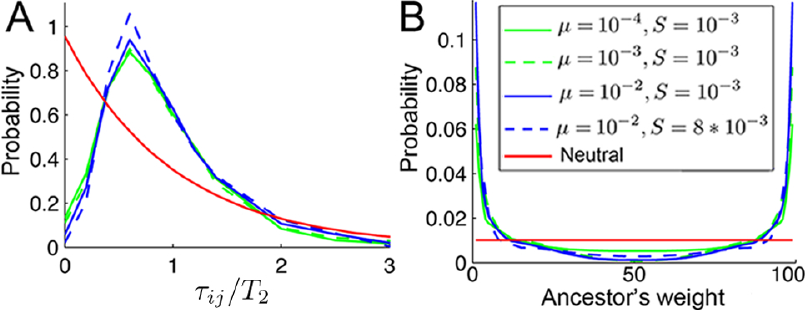}
 \caption{Distortion in the shape of genealogies in the presence of selection. (A) Distribution of pairwise coalescent time, scaled with its mean, $T_2$.  (B) Probability of an ancestor to carry  weight $w$ when there are $a=2$ lineages left in the genealogical tree of $n=100$ samples. Distributions based on 8000 random samples and population replicas. $N=64000$ and $\epsilon=0.1$ in both panels. \label{tree distort}} \end{figure}

One immediately notes two striking (and well known \cite{kirkpatrick1993searching,maia2004effect,seger2010gene} differences distinguishing Fig. \ref{fit_dis}D-E and F: Fitness diverse populations  i) have long terminal legs and are compressed towards the MRCA root of the tree, ii)  exhibit strong asymmetry of branching. These  anomalies are quantified in Fig. \ref{tree distort}.
Figure \ref{tree distort}A presents distributions of pairwise coalescent times in the population, $\tau_{ij}$, for  $\{i,j\}$ genome pairs for several parameter sets. In the Kingman's coalescent, $\tau_{ij}$ has an exponential distribution (with mean $N$) \cite{hein2005gene,wakeley2008coalescent} and most lineages in a genealogical tree coalesce at early times. In contrast, the bulk of coalescence in a population under selection is significantly delayed (compared to the total coalescent time) - an effect corresponding to the comb-like appearance of the trees.

The asymmetry of branching is quantified in Fig. \ref{tree distort}B which presents the distribution of weights at the level just below the MRCA, where there are only two ancestral lineages left in the tree. The strong bias toward extreme values of $w$ in populations under selection is to be contrasted with $w$-independent distribution predicted and observed in the neutral case (see SM). In SM, we consider more quantities reflecting the differences between the shape of trees. For example, we consider the probability distribution for the derived allele frequency $\nu$ and show that in the presence of selection it falls as $1/\nu^2$ with an upward bend for high frequencies, which is to be contrasted with the $1/\nu$ dependence in a neutral dynamics.

\subsection{Correlation between ancestral weight and offspring fitness}  
\begin{figure}[t]
\includegraphics{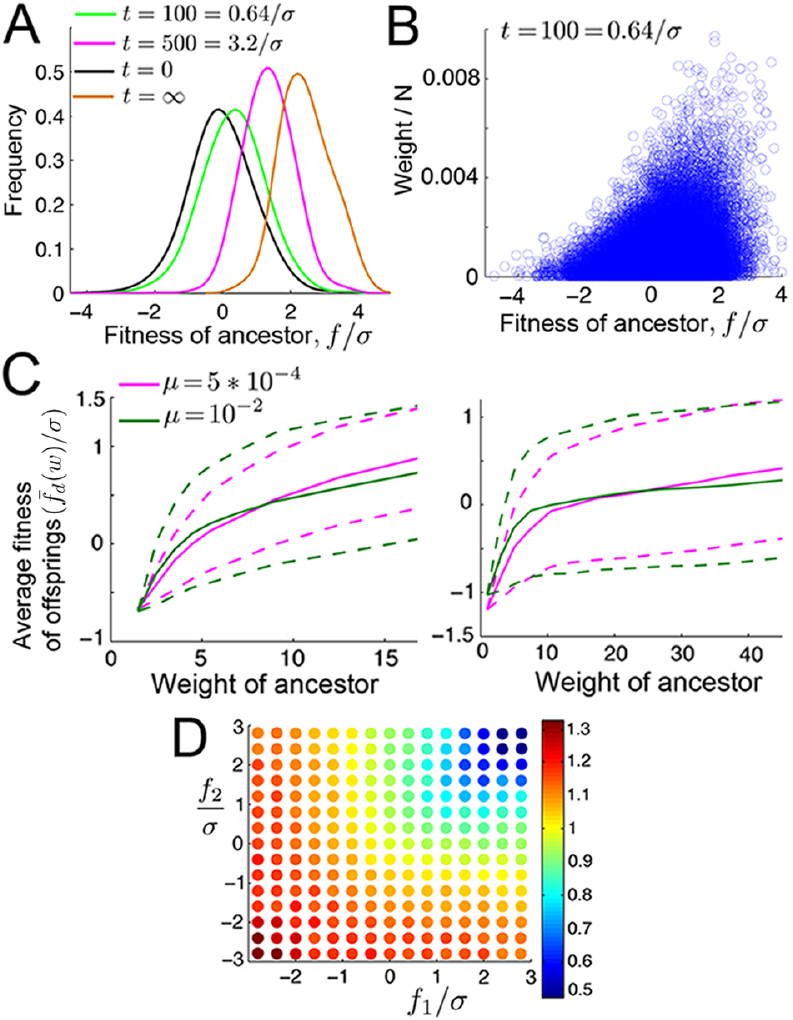}
 \caption{Correlation between fitness, weight and coalescent time. (A) Fitness distribution of the ancestors of the whole population, for a few time intervals in the past. Fitness is measured at the time the ancestor existed. Shown in the legend is also the time values in the unit of $1/\sigma$. $N=64000$, $\epsilon=0.1$ and $s=2*10^{-3}$ in all the panels.  $\mu=10^{-3}$ in (A), (B) and (D). (B) Scatter plot of  weight versus the fitness of the ancestors at $t=100$ generations ago. (C) Average fitness of offspring as a function of the ancestral weight in a sample of size $n=100$ at two different time slices in the past shown using solid lines. The dashed lines represents the standard deviation, $\bar{\Sigma}_d(w)/\sigma$, above and below the mean, $\bar{f}_d(w)/\sigma$. The two time points are chosen to be the first time that the tree carries a lineage with weight greater than $15\%$ and $40\%$ of  $n$.  Since the curves for various parameter sets were similar, for the sake of clarity, we only show them for two sets. (D) Heat map of mean pairwise coalescent time as a function of the fitness of the involved genomes, $f_1/\sigma$ and $f_2/\sigma$, normalized by the mean pairwise coalescent time for the whole population: $t_2(f_1,f_2)/T_2$. 
\label{anc_fit_time_seri} }\end{figure}

Let us consider the whole population and trace the surviving lineages  back in time, identifying all ancestors of the present day population $t$-generations in the past.  Figure \ref{anc_fit_time_seri}A shows the distribution of the ancestral fitness  (relative to the mean for that generation) at several time points in the past. This distribution becomes progressively shifted towards higher fitness as compared to the distribution for the whole population \cite{o2010continuous}.  In the limit of large times, this distribution follows the fitness dependence of the non-extinction probability of a lineage \cite{neher2010rate,neher2012genealogies}.

We shall be particularly interested in the time in the past when there is still a large number of ancestors, e.g. about $10^{3}$  at $t=100$.  Figure \ref{anc_fit_time_seri}B shows the scatter plot of the weight of ancestors versus their fitness advantage. Note that, by collapsing the points on the fitness axis, one gets the histogram shown in Fig. \ref{anc_fit_time_seri}A for $t=100$. We observe a strong positive correlation between the weight and the fitness of an ancestor. Higher fitness individuals in the past generations are not only more likely to survive, but they also leave more offspring conditional on the survival. Thus the weight of the ancestor, which can be determined from a reconstructed genealogical tree, can be used as a proxy for ancestral fitness: a quantity that one does not expect to know directly, except in the case of computer simulations! In SM, we provide plots of average ancestral fitness conditioned on its weight for various time points and parameter sets and confirm that the positive correlation between the weight and the fitness of ancestors holds quite generally. This correlation decreases as the time shifts further into the past.
  
\textit{Next we examine the correlation between the weight of the ancestor and the fitness of its surviving progeny. Consider a sample of genomes with size $n$ and the corresponding genealogical tree. One expects genomes which are derived from relatively high fitness ancestors to belong to higher fitness classes in the present time. Since ancestral fitness correlates with weight, we expect higher weight ancestors to produce, on the average, higher fitness descendants. To see this, let us consider an ancestor $i$ that existed some $t$-generations in the past. Assume this ancestor is carrying some weight $w_i$. We examine the fitness $\{f_{1},...,f_{w_i}\}$ of the $w_i$ offspring in the sample stemming from that ancestor. In particular, we focus on the mean, $f_d(w_i)=\frac{1}{w_i}\sum\limits_{j=1}^{w_i} f_{j}$,  and the variance, $\Sigma_d^2(w_i)={\frac{1}{w_i}\sum\limits_{j=1}^{w_i}( f_{j}-f_d(w_i))^2}$ over the $w_i$ offspring (subscript $d$ refers to descendants). Let us denoted the average of these quantities over random samples of genomes and over population replicas by $\bar{f}_d(w_i)=<\frac{1}{w_i}\sum\limits_{j=1}^{w_i} f_{j}>$ and $\bar{\Sigma}_d^2(w_i)=<\frac{1}{w_i}\sum\limits_{j=1}^{w_i}( f_{j}-f_d(w_i))^2>$. Note that $\bar{f}_d(w)$ and $\bar{\Sigma}_d(w)$ depend on the time, $t$. }
  
In Fig. \ref{anc_fit_time_seri}C and D, we show $\bar{f}_d(w)/\sigma$ and $\bar{\Sigma}_d(w)/\sigma$ averaged over different population realizations at two different time points in the past for trees with sample size $n=100$ (see SM for other parameter sets). In both cases, the mean fitness of the derived genomes is an increasing function of the weight of their ancestor. From Fig. \ref{anc_fit_time_seri}C, we also notice that the variance in the fitness of the derived genomes, $\bar{\Sigma}_d(w_i)$, is higher for higher mutation rates.  Consider a time closer to the root of a tree (e.g. right plot in Fig. \ref{anc_fit_time_seri}C), where a lineage can carry a significant portion of the sample size.  As expected, the value of $\bar{f}_d(w)$ for such high-weight ancestors is close to zero (remember that $f_i$ was defined relative to the population mean, so that the average of $f_i$ over the whole sample is zero). At the same time $\bar{\Sigma}_d(w)/\sigma \rightarrow 1$ for ancestors with $w$ approaching $n$. Interestingly, for the lineages which are still carrying a small weight at late in the coalescence process, the value of $\bar{f}_d(w)$ is clearly negative. 

High-fitness genomes typically merge first in a tree and form high-weight ancestors. This fact is seen in the distribution of the pairwise coalescent time, $\tau_{ij}$,  shown in Fig. \ref{tree distort}A. Averaging $\tau_{ij}$ over all  $\{i,j\}$ pairs of genomes in a population gives the mean coalescent time $T_2$. Now, consider the average of $\tau_{ij}$ conditioned on the fitness of the two genomes and denote it by $t_2(f_i,f_j)$. Fig. \ref{anc_fit_time_seri}D shows a heat map of $t_2(f_i,f_j)/T_2$. For two genomes both with high-fitness, the average coalescent time is shorter than $T_2$.  This is because they are likely to be relatively recent lineages emanating from the "nose" of the distribution \cite{desai2007beneficial}. In other words, the chance of sampling identical or similar sequences is greater for fitter samples than for less fit samples, since, fitter samples have shorter average pairwise coalescent time. This observation is the key to the proposed fitness inference method.

\subsection{Relative fitness inference based on the reconstructed genealogy}
Above we have reviewed the different ways in which the shape of the genealogical tree of the population under selection differs from a neutral one and have demonstrated the correlation between ancestral weights and the fitness of the corresponding descendants. We now show that this insight can be converted into a method for inferring relative fitness of genomes within the sample. 

To that end, let us consider a randomly chosen set of $n$ genomes from a population and use standard phylogenetic tree-building methods (see SM) to approximately reconstruct the genealogy of the sample. The accuracy of the reconstructed genealogy compared to the actual genealogy, known exactly from the forward simulation of population dynamics, is discussed in the SM. It increases with the neutral mutation rate $\mu_0$: in the biologically plausible regime of $\mu_0/\mu \approx 10$ considered here, it proves more than adequate to enable meaningful inference.

Next, based on the reconstructed tree, we associate with each leaf $i=1,...,n$ a {\it fitness-proxy score} (FPS), $\phi_i$,  defined by its lineage within the tree. Specifically, we define $\phi_i$ as a linear discriminator in the form
\begin{equation}
\phi_i = \sum_{k=1}^{m_i} \Theta (t_{a_k(i)}/T_2) [ w_{a_k(i)}-w_{a_{k-1}(i)} ] 
\end{equation}
where $\{a_k(i)\}$ is the lineage of genome $i$, starting with the genome itself as $a_0 (i)$ and running the length, $m_i$, of the lineage (i.e. the number of nodes)  until the root of the tree.  When an ancestral lineage $a_{k-1}$ merges with an internal node $k$, it forms a new ancestral lineage $a_{k}$ \textit{(see SM for further clarification of the notation using an example of a tree)}. The time of formation of the corresponding internal node is denoted by $t_{a_k(i)}$. The parameter $T_2$ is the estimate of the average pairwise coalescent time, obtained from the sampled genomes. Finally,  $\Theta(x)$ is a "soft step" function (a.k.a. Fermi function): $ \Theta(x) = \big(1+\exp(\beta ( x/ x_* -1 ))\big)^{-1}$ 
parametrized by the position of the step $x_*$ and its characteristic width $\beta$. If the $\beta \gg 1$ function $\Theta (x)$ steps abruptly from one to zero as $x>x_*$, so that $\phi_i = w_{a_*}-1$ where $a_*$ is the oldest ancestor in the lineage with $t_{a_*} < x_*T_2$. For $\beta \sim 1$ the FPS is defined by a weighted sum of ancestral weights from the $t_{a} \sim  x_*T_2$ "era" (see SM for details).

The logic behind our heuristic choice of the specific form of $\phi_i$ is to exploit the correlation between the offspring fitness and ancestral weights, which, at least on the high fitness/ high weight end of the distribution, decreases for $t_a > T_2$, because at long times even the lineages originating from high fitness ancestors spread all over the surviving population. Hence we choose $x_*<1$:  specifically the results below were obtained with $x_*=0.5$ and $\beta =5$, but in the SM we examine the performance of the ranking algorithm as a function of the parameters and demonstrate that nearly optimal performance (at least for the present form of the FPS) is achieved for a broad range of $x_*, \beta$. Critically, normalization of $t_a$ to the characteristic time of coalescence for the sample,  $T_2$, essentially eliminates the need  to know the evolutionary parameters of the population, such as its effective $\mu/s$ or $N$.

\begin{figure}[t]
\includegraphics{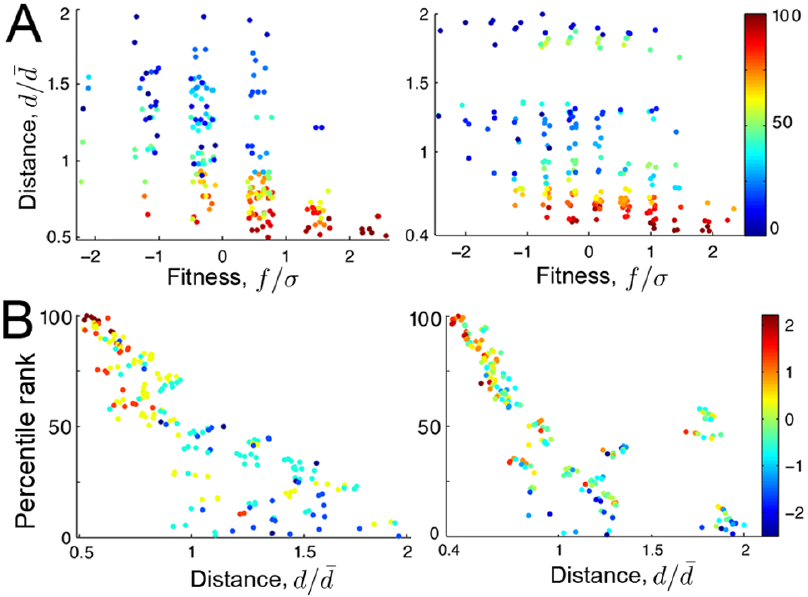}
 \caption{Examples of performance of the ranking algorithm. (A) Heat map of rank as a function of  fitness and  average distance to the fittest 10\%  genomes. Distance $d$ is normalized by its mean  $\bar{d}$. Left and right panels correspond to two samples of size $n=200$ drawn from the same populations as Fig. \ref{fit_dis}A ($\mu=10^{-3}$) and Fig. \ref{fit_dis}B ($\mu=10^{-2}$), respectively.  To avoid overlap of points, a small random number has been added to the fitness coordinate of each.  (B) Scatter plot of rank versus  distance to the top 10\% fittest genomes (colormap represents $f/\sigma$). The panels correspond to the same trees as in (A).\label{score}} \end{figure}

We rank genomes according to their $\phi_i$ score and compare this ranking with the actual fitness of each genome. In addition to inferring relative fitness, it is useful to know how genetically close a given genome is to the fittest in the sample.  Hence, for each genome we define $d_i$ as average of its Hamming distance to the fittest 10\% genomes in the sample. Figure \ref{score}A-B shows the results of the ranking for two $n=200$ samples from the populations that already appeared in Fig. \ref{fit_dis}A and B. We observe a strong correlation between FPS  ranking and the actual fitness in general and the "tendency" (quantified below) for the fittest genomes of the sample to show up in the top ranks. In addition, highly ranked genomes have smaller $d_i$ values indicating that they are genetically close to the fittest 10\%.  

\begin{figure}[t]
\includegraphics{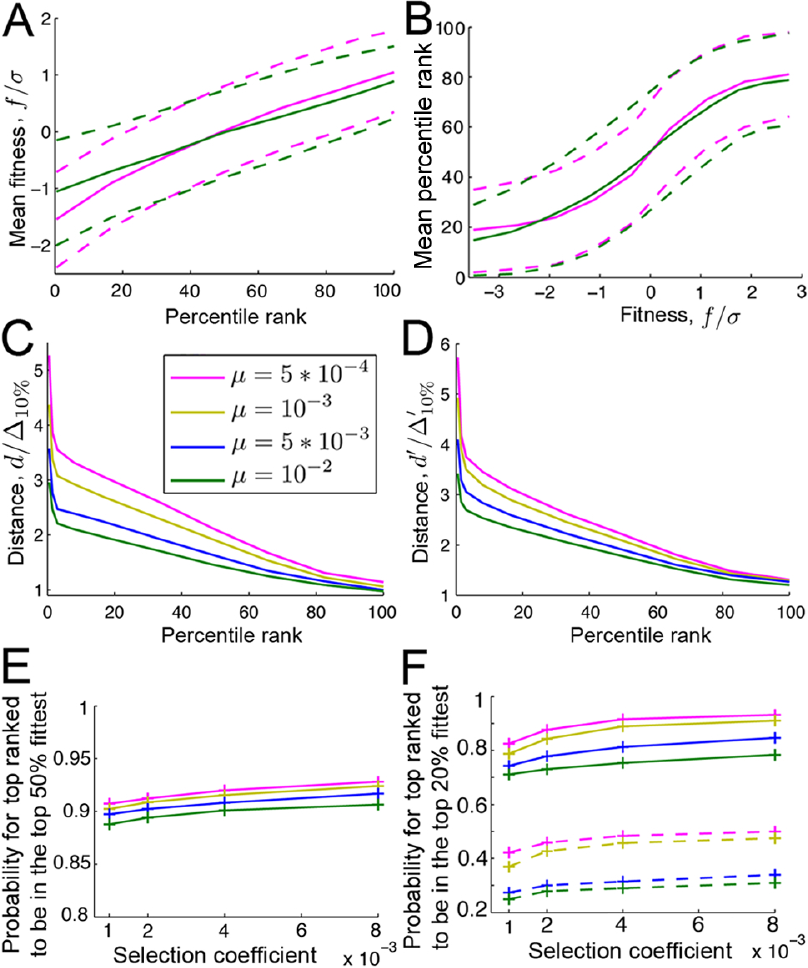}
 \caption{Performance of the fitness ranking algorithm. (A) {\it Solid lines}: mean fitness as a function of rank. {\it Dashed lines}: standard deviation above and below the mean ($\mu=5*10^{-4}$ and $10^{-2}$, see panel (C) for the legend). (B) Same as (A) for mean rank as a function of fitness.  (C)  Mean Hamming distance to the top 10\% fitness set, normalized by $\Delta_{10\%}$ (see  text) as a function of rank. (D)  Mean Hamming distance to ancestors of the generation at one turnover time in the future, normalized by $\Delta'_{10\%}$ (see  text) as a function of rank. (E) Probability for the fitness of a genome within the top \%10 ranked  to belong to the top $50\%$ fitness values  of sampled genomes for a range of mutation rates and selection coefficients. (F) Probability for the fitness of a genome within the top  10\% ranked to belong to the top 20\% fitness values shown using solid lines. The dashed lines show this probability for a randomly chosen genome (see the main text). Sample  size $n=200$, $N=64000$ and $\epsilon=0.1$ in all cases; $s=2*10^{-3}$ in (A-D).  
\label{top_50}} \end{figure}

The above observations are confirmed and quantified by the statistical data obtained by repeating the comparison for 8000 independent population samples and different sets of parameters. Specifically,  Fig. \ref{top_50}A shows mean fitness conditional on the FPS ranking and  Fig. \ref{top_50}B shows the mean rank conditional on actual fitness (normalized by $\sigma$) for two different values of $\mu$.  Figure \ref{top_50}C shows mean distance from the fittest conditional on the FPS ranking (for four different values of $\mu$), with distance normalized to  $\Delta_{10\%}$ defined as the average $d_i$ amongst the fittest 10\%.  Remarkably, we observe that  $d/\Delta_{10\%}$ for the highest ranked genomes gets close to one, indicating good convergence, in the sense of Hamming distance, of the top ranked genomes to the fittest set. Further analysis of the algorithm's performance, as well as additional parameter sets including the case of $\epsilon =0$, can be found in SM.

Looking at Fig. \ref{score}A we note greater dispersion in the fitness of the highly ranked subset, then in the distance. Indeed, some genomes which are not among the fittest can still be genetically close to the fittest subset: e.g. note in Fig. \ref{fit_dis}A-B the genomes with blue color located close to the mostly orange/red clusters on the right side of the trees. This is because the Hamming distance is dominated by neutral mutations $\mu_0 \gg \mu$, and is less susceptible to fluctuations compared to fitness, which is defined by a much smaller number of non-neutral mutations. To the extent that genetic relatedness is defined by the distance, the latter is essential for identifying within the sample the closest relatives of future populations. Taking advantage of ready accessibility of evolutionary future within our simulations, we have directly tested the ability of our approach to identify, within the sample, the genotypes that are closer to future populations. For each sampled genome, we define $d_i^\prime$ as the average of its Hamming distance to all of the genomes in the current population that are direct ancestors of the population in a generation about one genetic turnover time in the future (we know these ancestors from the forward simulation).  Typically, less than 100 individuals from the current population of $N=64000$ have descendants in this future population. In each case we normalized the distances by $\Delta'_{10\%}$ defined as the average of the smallest 10\% values of $d'_i$. Figure \ref{top_50}D shows $d_i^\prime/\Delta'_{10\%}$ conditional on the FPS ranking. We again observe that  $d'/\Delta'_{10\%}$ for the highest ranked genomes gets close to one, indicating that the top ranked genomes are indeed  close to the ancestors of future generations. This means that the FPS ranking makes possible to identify the genetic elements (common among the high rank genomes) that future populations inherit from the present one.

Finally, we examine the fitness of the genomes with the 10\% highest rank. Consider the sorted vector $F=[f_1,...,f_n]$ which contains the actual fitness values for all the sampled genomes. In Fig. \ref{top_50}E, we show that the probability for the fitness of a genome within the top 10\% rank to be above the median fitness is about $0.9$, for the broad range of parameters considered. The probability for the fitness of a top 10\% -ranked genome to belong to the top $20\%$ fitness class is given by the solid lines in Fig. \ref{top_50}F and is above 0.7. Note that some of the sampled genomes can have equal fitness (i.e. $F$ contains duplicate values), which is more common for lower mutation rates where the fitness diversity in the populations is limited. Hence, to provide a meaningful comparison for this probability, in Fig. \ref{top_50}F we show - dashed lines - the probability for a random genome to be in the top $20\%$ fittest. Clearly, the top $10\%$ ranking is a good predictor of high fitness.

In summary, above results clearly indicate the power of the proposed inference method. The performance of the method improves monotonically with the increasing sample size (see SM):  it degrades significantly, compared to the results presented above, for $n<100$ but approaches saturation for $n>200$. 

\section{Discussion} 
Whereas one often thinks of evolution occurring on geological time scales, evolutionary dynamics can also unfold swiftly as it does in bacteria acquiring antibiotic resistance, in HIV evading CTL response in the course of infection or in the progression of an aggressive cancer. Recent advances in sequencing \cite{smith2010highly,navin2011tumour} have made it possible to extensively sample such rapidly evolving populations. The amount and quality of genomic data on populations will only continue to increase, accentuating the challenge of extracting more information from sampled genomes. Here, we have demonstrated that the shape of genealogical trees contains much more  information than merely the evidence for (or against) selection within population. As a proof-of-principle we have  formulated a method for estimating relative fitness of individual genomes sampled from a fitness diverse but otherwise unstructured population, in the absence of any information other than genomic sequence. This provides the possibility of forecasting the common genotype of the future on the time scale of genetic turnover.

Our demonstration was based on  a vast simplification of biological and ecological reality. Our  model assumed fixed population size and constant environment; it neglected epistasis and assumed all non-neutral mutations (both deleterious and beneficial) to have the same effect size. While we have, within the model considered, explored a biologically interesting range of parameters, it would be useful to extend the study to a broader class of models. 
Yet, we expect the proposed method to be quite robust, because it is based on the very fundamental aspect of evolutionary dynamics, realized when population and the mutation rate are sufficiently large to harbor substantial non-neutral diversity, and when fitness differentials between individuals are formed by the contributions of numerous weakly selected loci rather than a small number of strong ones. In this multi-locus weak selection regime, surviving lineages in the course of time move from the nose of the fitness distribution towards the center, in an biased diffusion fashion. The correlation between  early coalescence and rapid increase of ancestral weight along the lineages with high relative fitness, derives from the continuous genetic turnover of the population described above. This turnover occurs in traveling waves models corresponding to the continuous adaptation scenario  \cite{tsimring1996rna,rouzine2003solitary}, in the dynamic mutation-selection balance \cite{goyal2011rare} which involves both deleterious and compensating beneficial mutations, as well as in the case of  purifying selection ($\epsilon=0$) \cite{gordo2000degeneration,walczak2012structure}.

A detailed statistical analysis of the way lineages propagate along the fitness axis could allow to improve FPS by optimizing the tradeoff between gaining more information about a particular lineage by tracking it further back in time and the loss of predictive power due to the fact that beyond the genetic turnover time even lineages of the fittest ancestors spread all over the fitness distribution. Presently, we have dealt with the problem heuristically by focusing on the coalescence sequence for each lineage up to about $0.5 T_2$. The advantage of our simple heuristic approach is that it is more likely to be model independent than the more fine-tuned methods.

It would be interesting to extend the fitness inference method to recombining populations. This should be relatively straight-forward as long as genetic turnover time is faster compared to the inverse recombination rate. For a chromosome with an approximately uniform crossover probability, this condition defines a characteristic length below which loci coalesce in essentially recombination-free genealogies.  Roughly, the asexual coalescent considerations would apply to a 1cM size locus provided that it harbors $\sigma > 10^{-2}$.  More careful analysis is however necessary in order to deal with the Hill-Robertson effect or {\it genetic draft} \cite{hill-roberston,neher2011genetic} caused by the transient linkage of the locus to the rest of the genome which effectively adds noise, reducing effectiveness of selection on the individual loci.

Clearly, the highest priority for the future would be to test the method on experimental or epidemiological data.  Applications are possible wherever genomic data is available for fitness-diverse, but otherwise unstructured populations.
Genomic data from single cell sequencing of tumors \cite{navin2011tumour} or from localized influenza outbreaks \cite{squires2011influenza} are among the interested possibilities to be considered. For example, it would be interesting to compare the proposed method with the clustering-based approach of \cite{plotkin2002hemagglutinin} to predicting antigenic evolution of influenza A. In addition to predicting which genotypes are more likely to appear in future generations, fitness inference method could be used for QTL mapping \cite{broman2009guide} with FPS-based ranking being the quantitative phenotype that could be used to identify highly adaptive or deleterious alleles.

\section{Acknowledgements}
We thank Richard Neher, Daniel Balick and Sidhartha Goyal for many useful discussions. AD was supported by HFSP RFG0045/2010 and NSF PHY11-25915  while BIS acknowledges support of NIGMS R01 GM086793.

\section*{Appendix A:  Supplemental materials}
\subsection{Example of a Tree to Describe the Notation}
\begin{figure}[btp]
\includegraphics{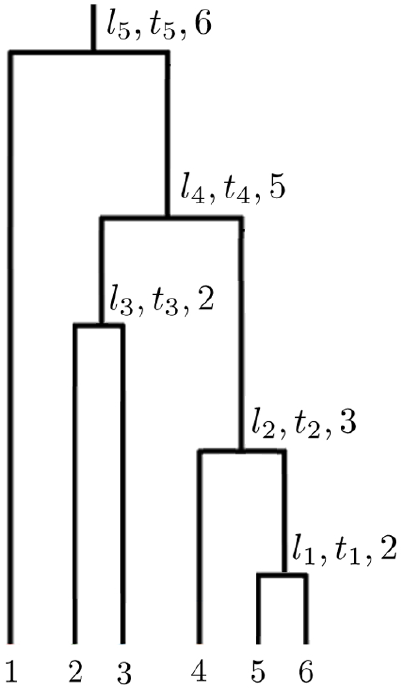}
\caption{Example of a tree formed by 6 individuals. Each of the starting 6 lineages carry weight 1 and have fitness $f_1$ to $f_6$. Lineage $l_1$ is formed at time $t_1$ by the merger of two lineages 5 and 6, and carries weight 2. Similarly, lineage $l_2$ is formed at time $t_2$ by the merger of two lineages 4 and $l_1$, and carries weight 3. \label{example_tree} } \end{figure}

\textit{Consider the tree shown in Fig. \ref{example_tree} from a sample of $n=6$ individuals. The figure shows the label, the time of formation and the weight of each ancestor associated with each internal node. In the paper, we have referred to quantities such as the mean ($f_d(w_i)$) and the variance ($\bar{\Sigma}^2_d(w_i)$) in the fitness of the offspring of an ancestor existing some $t$-generations in the past. As an example, in Fig. \ref{example_tree}, consider some time $t$ in the past such that $t_3<t<t_4$, where three lineages exist: $1$, $l_2$ and $l_3$. The lineage $l_3$ carries weight 2 with descendants 2 and 3 in the sample. In this case, the mean and the variance in the fitness of the offspring of lineage $l_2$ is given by $f_d(3)=\frac{1}{3}(f_4+f_5+f_6)$ and $\Sigma^2_d(3)=\frac{1}{3}((f_4-f_d(3))^2+(f_5-f_d(3))^2+(f_6-f_d(3))^2)$, respectively. Similarly, for lineage $l_3$, the two quantities are given by $f_d(2)=\frac{1}{2}(f_2+f_3)$ and $\Sigma^2_d(2)=\frac{1}{2}((f_2-f_d(2))^2+(f_3-f_d(2))^2)$. Finally, for lineage 1 we get $f_d(1)=f_1$ and $\Sigma^2_d(1)=0$.One can repeat a similar procedure for any other time point along the tree.}
 
\textit{We now explain the notation used in the formula for Fitness Proxy Score. First we need to calculate $T_2$, the estimate of the average pairwise coalescent time, obtained from the sampled genomes. For the tree in Fig. \ref{example_tree}, the pairwise coalescent time between lineages 3 and 4 is $\tau_{3,4}=t_4$, or between lineages 1 and 4 is $\tau_{1,4}=t_5$. One gets $T_2=\frac{1}{15}(t_1+2*t_2+t_3+6*t_4+5*t_5)$. As an example, let us calculate the score of individual 3, $\phi_3$, in the tree presented in Fig. \ref{example_tree}. There are $m_3=4$ ancestral lineages to individual 3: i) the lineage 1 starting from the individual 3 itself, ii) $l_3$, iii) $l_4$ and iv) $l_5$. These four lineages are denoted by $a_0$ to $a_4$, respectively.  At time $t_3$, the lineage $a_0$ with weight 1 merges with another lineage with weight 1 and form $a_1$ with weight 2. The contribution of this coalescent event to $\phi_3$ is equal to $\Theta (t_3/T_2)*(2-1)=\Theta (t_3/T_2)*1$. Proceeding in a similar fashion, we obtain $\phi_3=\Theta (t_3/T_2)*1+\Theta (t_4/T_2)*3+\Theta (t_5/T_2)*1$. Note that if we ignore the time dependent coefficients $\Theta (t_i/T_2)$, all the individuals get the same score of $n-1=5$.}

\subsection{Evolutionary simulations}
The simulations are done using a custom written Python code, available upon request. The evolution is based on a discrete time Wright-Fisher model with population size $N$. Each generation $t$ undergoes  separate selection and mutation steps. To implement selection, each individual $i$ produces a Poisson-distributed number of gametes in the next generation with parameter $\exp(f_i-\alpha)$. Here $f_i= F_i - \bar{F}$  is the fitness advantage of individual $i$ relative to the mean fitness of the population $\bar{F}$, and $\alpha = {\frac{N(t)-N}{N}}$ ensures an approximately constant population size around $N$. Individual genomes are defined as binary strings, $g_k$ with $k=1,...,L$ and the number of loci, $L=10^{5}$,  chosen large enough to exceed the number of segregating polymorphisms in the simulated population. Consistent with infinite site approximation, new mutations flip the $g_a$ binary value from zero to one.  

At each generation, the beneficial and deleterious mutations arise with probability $\epsilon \mu$ and $(1-\epsilon) \mu$ and have a fitness effect of $\pm s$, respectively. We also record the forward genealogies during the simulations. The above process is repeated for a specified number of generations. Various information on the dynamics of the evolution are measured after an equilibration time to remove transient effects from the initial conditions. In the parameter regimes studied, we found that $10^4$ generations was generally sufficient.  
 
Given that we perform forward simulations and keep track of the genealogies, the simulations are computationally intensive. Therefore, the maximum population size that we simulated was $N=64000$.  The mutation rate was varied from $\mu=10^{-4}$ to $\mu=10^{-2}$  and the selection coefficient from $s=10^{-3}$ to $\mu=8*10^{-3}$. For the parameter combination where $N=64000$, $\epsilon = 0.1$ and  $\mu=10^{-4}$  (beneficial mutation rate $10^{-5}$ and deleterious mutation rate $9*10^{-5}$), only a couple of clones are segregating in the population (see below). This parameter combination serves as the boundary between the multisite selection regime and the selective sweep regime. For smaller mutation rates, given that $N=64000$, the population is monoclonal and enters the regime of selective sweeps.

\begin{figure*}[btp]
\includegraphics{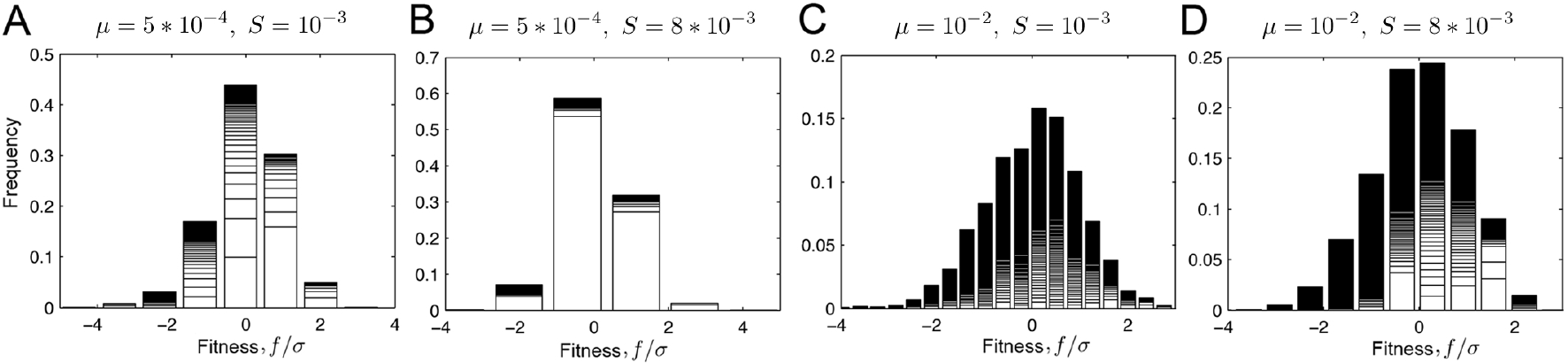}
 \caption{Fitness distribution at one time slice. The mutation rate and selection coefficient for each case is written on the top of the corresponding panel.  $N=64000$ and $\epsilon=0.1$ for all the panels.  Each bin corresponds to a fitness class and each class is composed of several clones. The height of each box within each bin represents the size of a clone. Larger clones are stacked on the bottom. The dark band on top of each bin correspond to small clones.\label{more profiles} } \end{figure*}

\begin{figure*}[btp]
\includegraphics{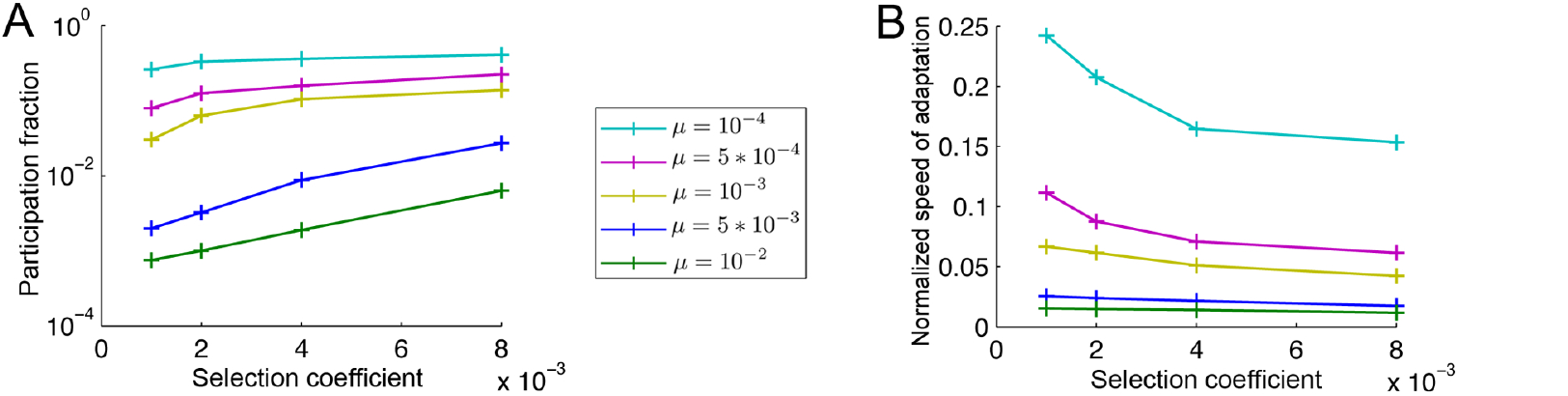}
 \caption{Clonal structure and adaptation rate. (A) The participation fraction, $Y$, for different parameter values. $Y$ is the probability that two randomly chosen genomes belong to the same clone. For all the curves, $N=64000$ and $\epsilon=0.1$. When $\mu=10^{-4}$ (beneficial mutation rate $10^{-5}$ and deleterious mutation rate $9*10^{-5}$), the values of $Y$ become significant ($>0.1$). This implies that the dynamics is at the boundary between the multisite selection regime and the selective sweep regime. (B) Speed of adaptation, normalized by its expectation value in the limit of selective sweep $2N\mu \epsilon s^2$. \label{partic}} \end{figure*}

Below, we present some results on the clonal diversity, as well as the speed of adaptation, for various parameters that we have simulated. In Fig. 2A and B of the main text, we showed two examples of fitness distribution. In Fig. \ref{more profiles}, we show some more examples. Assume there are $c$ clones in the population, with sizes $n_1,...,n_c$. Note that $\sum\limits_{i=1}^c n_i=N$. To see how many clones with significant size are segregating, we can define the participation fraction: $Y=<\sum\limits_{i=1}^c (\frac{n_i}{N})^2>$. This quantity is equal to the probability that two randomly chosen genomes belong to the same clone. Fig. \ref{partic}A shows the participation fraction ranging  from values smaller than 0.001 to values around 0.1 for various sets of parameters. For $N=64000$ and $\mu=10^{-4}$,  $Y\approx 0.4$, which means that for the smallest value considered for $\mu$, there is a significant probability that two randomly chosen genomes come from the same clone.

In the regime of our interest, where many mutations simultaneously segregate, it is well known that the competition between  mutations slows down the rate of the adaptation (Hill-Robertson or Fisher-Muller effect) \citep{desai2007beneficial,rouzine2008traveling}. In Fig. \ref{partic}B, we present the speed of the adaptation, i.e. the rate of change of the mean fitness, normalized by its expected value in the selective sweep regime. In the later regime, the beneficial mutations are rare enough that only a single mutation segregates at a time, and   assuming the deleterious mutations are purged, the expected speed of adaptation is $v=2N\mu \epsilon s^2$.   As we see in Fig. \ref{partic}B, for the parameter combination $N=64000$ and $\mu=10^{-4}$, the adaptation rate is only around a quarter of its expected value in the selective sweep regime. We see that the normalized speed of adaptation varies from 0.01 to 0.25 in the parameter range that we have considered. 

\begin{figure*}[btp]
\includegraphics{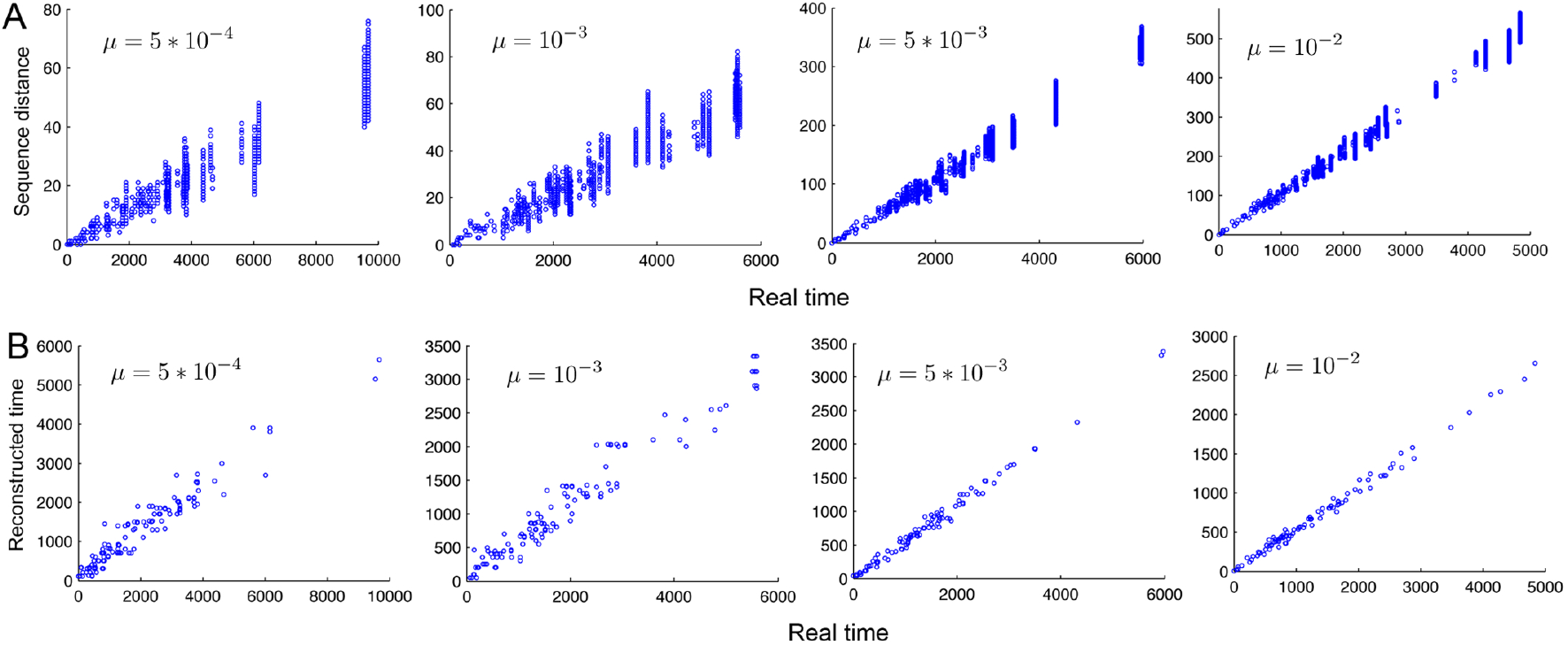}
 \caption{Examples of tree reconstruction for sample size of $n=100$. $N=64000$, $\epsilon=0.1$ and $s=2*10^{-3}$ in all cases. (A) Scatter plot of the Hamming distance between sequences versus the real divergence time for all the pairs in the sample.  The non-neutral mutation rate for each case is shown in the associated plot. The neutral mutation rate was set to 10 times the value of the non-neutral mutation rate. (B) Scatter plot of the reconstructed divergence time between sequences versus the real divergence time for all the pairs in the sample. Each plot is associated to the same tree as in panel (A) for the same mutation rate. \label{example of reconstruction} }\end{figure*}

\subsection{Tree reconstruction} \label{Tree reconstruction}
In the first step, we use the neighbor-joining algorithm \citep{durbin1998biological,felsenstein2004inferring} to reconstruct the tree topology. The input distance matrix for this algorithm is simply given by the pairwise difference of sequences (Hamming distance) including both neutral and non-neutral mutations. The time to the common ancestor of two individuals is proportional to the number of neutral genetic differences between them. For a real data set, one may use a more realistic substitution model to infer the divergence time between pairs of genomes. We have considered the values of the non-neutral mutation rate over a few orders of magnitude. The neutral mutation rate was always set to 10 times the value of the non-neutral mutation rate. We use the neighbor-joining algorithm only to infer the topology of the tree. We do not use the length of the edges that are calculated in this algorithm. The reason is that we want all the leaves of the tree to be located at the current time and have the same distance to the root. 

In the next step we find the root of the tree based on the parsimony method. Each  point on the tree divides the sample into two groups. The root should be located at a point where the similarity between the two groups is minimal. We count the number of mutations which exist in both groups and assign the root to a point where this number is minimal.

In the last step, we assign the height (time interval to the present time) of each node in the tree. The lengths are calculated as in the UPGMA algorithm \citep{felsenstein2004inferring}. In this algorithm, the total branch length from a tip down to any node is half of the average of the distance between all the pairs of genomes whose most recent common ancestor is that node. We consider a node only after all the nodes below it have their heights assigned. We start from the bottom, namely, the nodes which are connected to two leaves. The height of these nodes are calculated similar to the UPGMA algorithm: the height is equal to the half of the mutational distance between the pair of the genomes below that node. For other internal nodes, we also calculate the putative height as half of the distance between all the pairs whose most recent common ancestor is that node. The height of the node is the maximum between this putative distance and the height of all the internal nodes below the considered node. 

We evaluated the performance of the above tree reconstruction algorithm in all different parameter ranges by comparing the reconstructed tree with the actual genealogy. In all the cases, the performance was satisfactory. In Fig. \ref{example of reconstruction}, we show examples of the performance of the above algorithm for four different mutation rates. For each rate, a sample of size $n=100$ is selected. In Fig. \ref{example of reconstruction}A, we show the sequence distance for all the $(n-1)n/2$ pairs in the sample versus the real divergence time. These distances are the input of the above algorithm. In Fig. \ref{example of reconstruction}B, we show the reconstructed divergence time (inferred from the reconstructed tree) for all the pairs. The validity of the above algorithm is reflected in the fact that the relation between these two times seems to be linear. The slope of the line is irrelevant, since, it only reflects an scaling factor, i.e. the estimation of the mutation rate. As we see in Fig. \ref{example of reconstruction}, for higher mutation rates (e.g. $\mu=5*10^{-3}$ and $\mu=10^{-2}$) where there is tens-hundreds of  differences between a typical pair of genomes,  the assumption of the neutral mutation rate being 10 time that of $\mu$ is unnecessary. In these cases, there is enough diversity that even setting the neutral mutation rate equal to $\mu$ would be sufficient.

\subsection{Weight Distribution} \label{Weight Distribution}
Consider a sample of size $n$ and the corresponding phylogenetic tree. Assume looking at the tree at the stage where there are $a$ lineages left. The ancestor $i$ will carry a weight $w_i$ where $i=1,...,a$ and $\sum\limits_{i=1}^{a}w_i=n$. The values that $w_i$ can take is anything between 1 and $n-a+1$. For example, when there are only 2 ancestral lineages, $w_i$ can be between 1 and $n-1$. The statistics of the phylogenetic trees for neutral evolution are given by the Kingman's coalescent \citep{kingman1982coalescent,kingman1982genealogy}. In particular, the probability distribution of $w_i$ is given by \citep{derrida1991evolution}:
\begin{eqnarray} \label{probab w}
P_{neu}(w_i|a,n)=\binom{n-w_i-1}{a-2}\ \Big/\  \binom{n-1}{a-1} 
\end{eqnarray}
For example, when there is only $a=2$ ancestors left in the tree, we get $P_{neu}(w|2,n)=\frac{1}{n-1}$, which is independent of $w$. In other words, when there are two ancestors left, each one can carry any weight between 1 to $n-1$ equally likely. The above formula can be derived solely based on the fact that, as one goes up in the tree, at each stage, any lineage is equally likely to coalesce with any other lineage regardless of the weight they are carrying or any other previous events in the tree. In the presence of selection, this is no longer the case and not all lineages are equally likely to coalesce. The probability of the coalescent between two lineages will depend on the history of previous merging events. 

\subsection{Distortion in shape of genealogical trees}\label{Tree Imbalance}
Here, we consider some quantities which  reflect the differences between the shape of trees from non-neutral and neutral evolution. While inspecting trees in Fig. 2D and E, we notice that in the presence of selection it is more common for a leaf (sampled genome) to be connected to a long edge. In other words, it takes a long time for some leaves to merge to other lineages in the tree. Moreover, such leaves are more likely to belong to lower fitness classes, represented by blue and grey colors.  In addition, number of lineages left in a tree as a function of time seems to be different. Here, we explore such points in more details. 

\begin{figure*}[btp]
\includegraphics{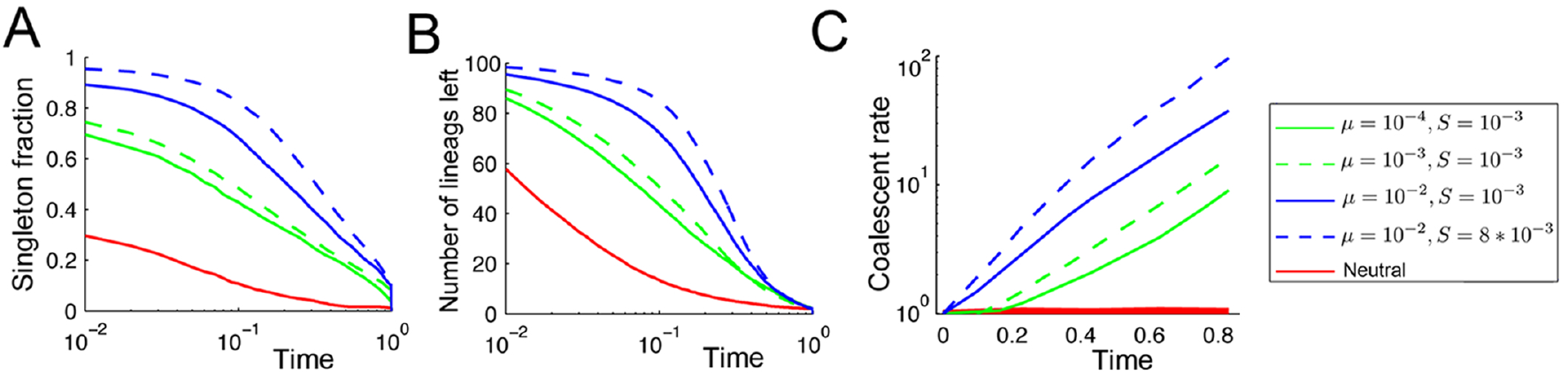}
 \caption{Statistics on lineages in phylogenetic trees. For all the panels, the sample size is $n=100$, $N=64000$ and $\epsilon=0.1$. (A) Average fraction of singleton lineages left in a tree as a function of time. The time for each tree has been linearly rescaled so that the root is at $t=1$ and the current time is 0.  (B) Average number of lineages left in a tree as a function of time. The time has been linearly rescaled as in part (A). (C) Coalescent rate between two random lineages as a function of time. The rate is normalized by its value at time $t=0$. The `effective population size', $N_e$, would be defined to be inversely proportional to coalescent rate. \label{sing_lin_coalrate}} \end{figure*}

At each instant of the time in a tree, one can consider what fraction of the remaining lineages are singletons. Singletons are defined as lineages with weight \textit{w=1}. In Fig. \ref{sing_lin_coalrate}A shows the average value of this fraction as a function of time. These curves are obtained by averaging over random samples and over population replicas. The time for each tree is linearly rescaled so that the current time is at 0 and the root is at 1. At time 0, all the lineages in a tree are singletons and the fraction is, therefore, one. At time 1, all the lineages have merged together and therefore no singleton lineage is left. As we see, the curve for the neutral case falls below the rest of the curves. One can ask, by looking at the number of singletons as a function of time for a single tree, is it possible to tell whether or not this tree is from a neutral population? Later, we show that the separation between the neutral and non-neutral curves is large enough that one can differentiate with high confidence whether or not a single tree is neutral.

In the neutral case, the statistic of length of singleton edges was studied in \citep{fu1993statistical} and is used in the Fu and Li's test for detecting departures from the Kingman's coalescent. As long as the sample size is not very small (e.g. $>5$), the expected total height of a neutral tree is nearly equal to $2N$ where $N$ is the population size. This quantity is almost independent of the sample size. The expected length of a singleton edge is given by $\frac{2N}{n}$ where $n$ is the sample size \citep{fu1993statistical}. In our example of tree in Fig. 2 of the main text where $n=30$, the average length of singleton edges is around $1/30$ of the total height of the tree. This means that most of such edges should be much shorter compared to the total height of the corresponding tree. However,  in the presence of selection, it is more common that some singletons survive even until close to the root of the tree. 

\begin{figure}[btp]
\includegraphics{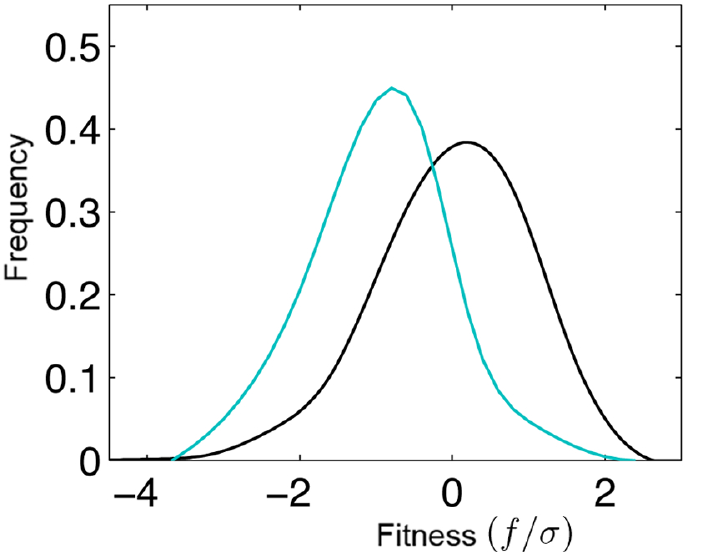}
 \caption{Fitness distribution of the singleton lineage which connects to the tree the latest is shown in cyan color. The black curve presents the fitness distribution of the whole population which is the same as fitness distribution of the sampled genomes. The distributions are obtained by averaging over random samples and over population replicas. $N=64000$, $\mu=10^{-3}$, $s=2*10^{-3}$  and $\epsilon=0.1$.  \label{singleton_fit}} \end{figure}

One can also look at the fitness of the singleton lineage that is the latest to join the rest of the tree. Fig. \ref{singleton_fit} shows the fitness distribution using the same simulation parameters as in Fig. 2D. As we see, these lineages tend to belong to the unfit classes. This is a general pattern observed for all of  the simulation parameters. 
 
We have also considered the average number of lineages left in a tree as a function of time,  $<a>_t$. The result is presented in Fig. \ref{sing_lin_coalrate}B. In the presence of selection, the number of lineages drops slower at early times compared to the neutral case. 
 Under neutrality, when there are $a$ lineages, the rate at which the next coalescent event happens is $\binom{a}{2}/N$ \citep{wakeley2008coalescent}. This is the product of the coalescent rate between two random lineages, $1/N$, and the total number of pairs among $a$ lineages, $\binom{a}{2}$. Therefore, the expected time a tree spends having $a$ lineages is $N/\binom{a}{2}$. In this case, the coalescent events happen much faster on the bottom of the tree, where $a$ is large, and most of the time in the tree is spent while having only a few lineages (also see Fig. 2F of the main text). On the other hand, in the presence of selection, coalescence times near the root are reduced compared to the total height of the tree. Alternatively, the external branches are longer compared to neutral expectations.

Under neutral evolution, since the coalescent events happen at rate $\binom{a}{2}/N$, when there are $a$ lineages left, one has: $\frac{d<a>_t}{dt}\propto-<\binom{a}{2}/N>$. Therefore,  the ratio $-\frac{d<a>_t}{dt}/<\binom{a}{2}/N>$ which is the coalescent rate between two random lineages remains constant. Fig. \ref{sing_lin_coalrate}C shows the coalescent rate normalized by its value at time $t=0$. For the neutral case, this rate remain constant, as expected. However, in the presence of selection, the rate increases for further time back in the tree. The reason for this is that, for times further back in the tree, the ancestral lineages are more likely to have belonged to the leading edge of the fitness distribution at the time they existed (see Fig. 4A of the main text). Therefore, they coalesce at a faster rate compared to the bottom of the tree where lineages are spread over the fitness distribution \citep{o2010continuous}. 

\begin{figure}[btp]
\includegraphics{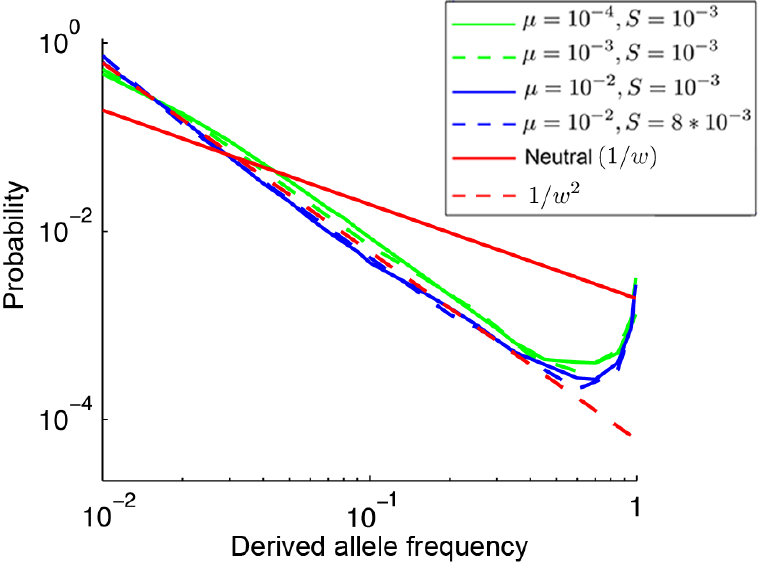}
 \caption{Distribution of neutral polymorphism in a sample of size $n=100$. The dashed red line shows the probability distribution which dependents on $1/w^2$, as opposed to $1/w$ (neutral case). $N=64000$, $\epsilon=0.1$. \label{freq} } \end{figure}

Increase in the coalescent rate is sometimes interpreted as a reduction in the effective population size (denoted by $N_e$). However, not all aspect of the coalescent process under selection, such as the weight distribution or fraction of singleton lineages, can be accounted for by only introducing an effective population size. This fact also manifests itself in the distribution of polymorphisms in a sample of genomes. Under neutrality (in the limit of infinite-site model), the probability that a derived allele appears in $w$ individuals out of $n$ sampled genomes is proportional to $1/w$. This behavior is a consequence of both the weight distribution and the length of coalescent intervals. To see this, note that in order to appear in $w$  genomes in the sample,  a mutation must have occurred on an ancestor with weight $w$. Assume this ancestor existed when there was $a$ lineages in the tree. The probability that 1 of the $a$ ancestors carried weight $w$ is $a*P_{neu}(w|a,n)$. The average time a tree spends having $a$ lineages is proportional to $1/\binom{a}{2}$. Summing over all possible $a$'s gives: $\sum_{a=2}^{n}\,  a P_{neu}(w|a,n) \frac{1}{\binom{a}{2}}=\frac{2}{w}$, which is the usual one over frequency dependence. In Fig. \ref{freq} we show the frequency distribution of neutral polymorphisms in the presence of selection. The distribution first drops more like $1/w^2$ for small frequencies and then bends upward for higher frequencies where $w > n/2$ \citep{neher2011genetic}.

\subsection*{\textbf{\textit{Test of neutrality based on the shape of trees}}}
Now, we discuss some measures for distinguishing between neutral and non-neutral trees. Let us use the term topology to refer solely to the branching pattern of a tree. On the other hand, the term shape will refer to the information about both the branching pattern and the branch lengths. In \cite{kirkpatrick1993searching}, authors reviewed six measures of tree asymmetry based solely on the tree topology. They studied the power of these measures to be used as a test for deviation of trees from neutral predictions. A similar analysis was carried out in \cite{maia2004effect}. One of the measures  denoted by $\sigma^2_n$ turned out to be relatively more powerful in both studies. Below, we show the result of applying this measure to the trees from our simulations. 

\begin{figure*}[btp]
\includegraphics{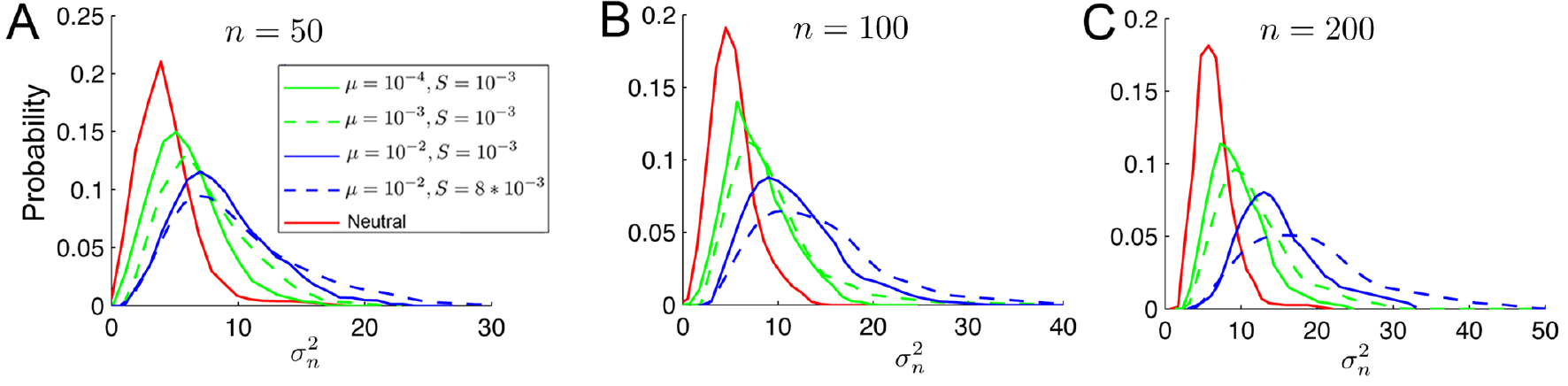}
 \caption{Distribution of $\sigma^2_n$, a measure of asymmetry based on the topology, for three samples sizes of $n=50$, $n=100$ and $n=150$. In all cases, $N=64000$ and $\epsilon=0.1$. \label{test sigma}} \end{figure*}

\begin{figure*}[btp]
\includegraphics{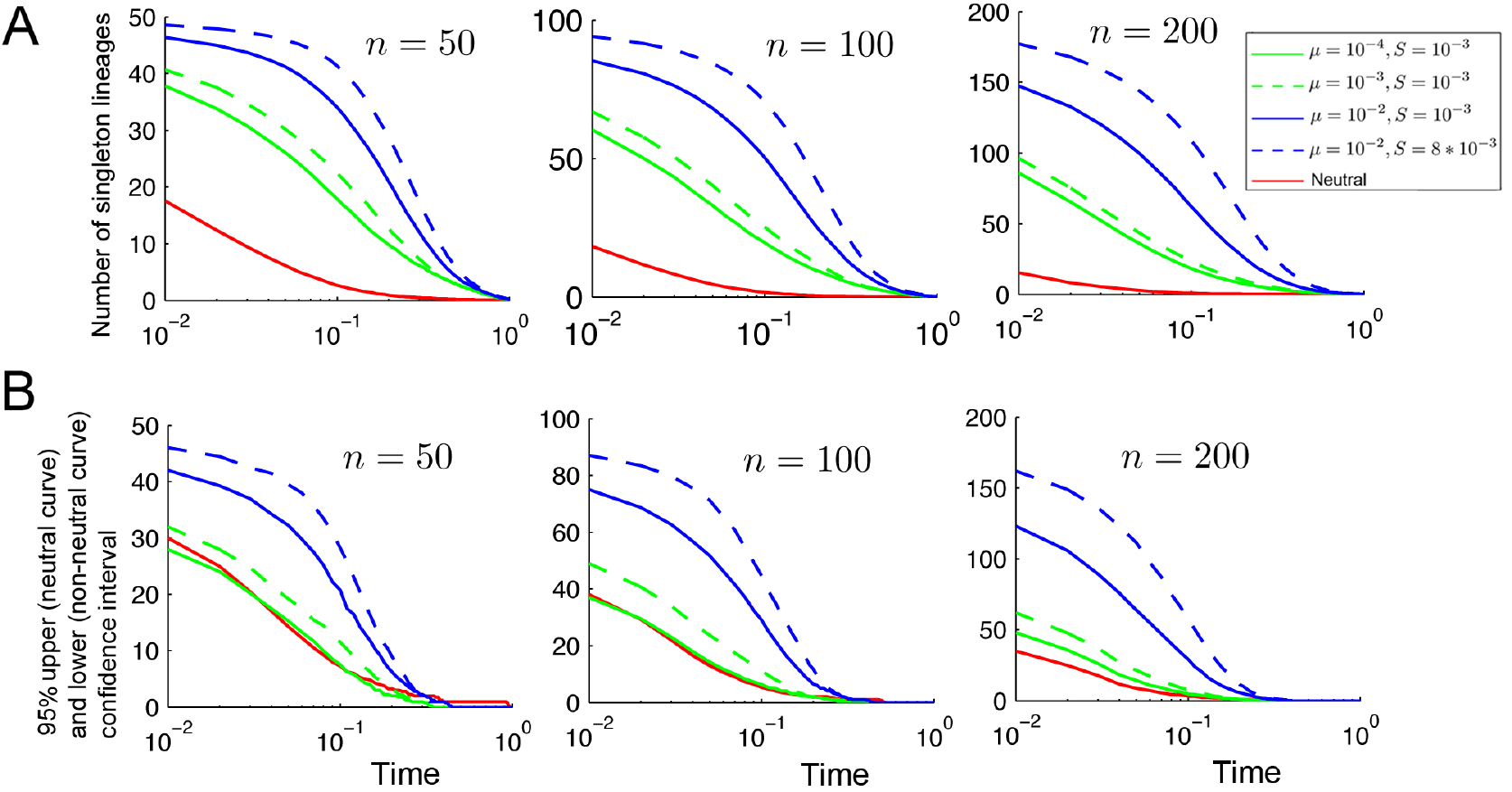}
 \caption{(A) The average number of singleton lineages left in a tree as a function of time, for three samples sizes of $n=50$, $n=100$ and $n=150$. The time has been linearly rescaled so that the root is at $t=1$ and the current time is 0. (B) The upper (for the neutral curve) and lower (for the non-neutral curves) $\%95$ confidence intervals for the curves in part A. In all cases, $N=64000$, $\epsilon=0.1$.  \label{test sing}} \end{figure*}
 
 \newpage
\begin{figure*}[btp]
\includegraphics{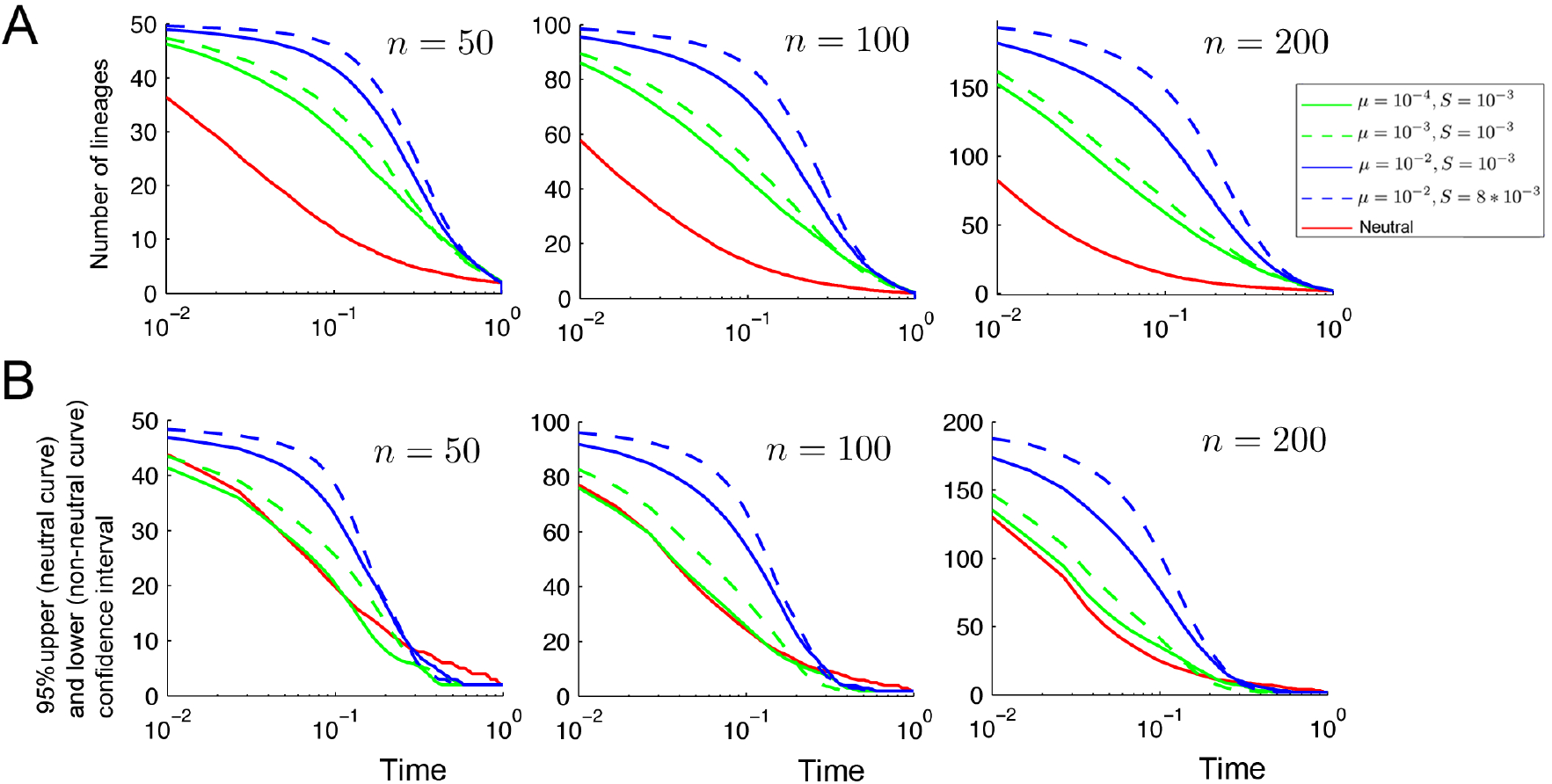}
 \caption{(A) The average number of lineages left in a tree as a function of time, for three samples sizes of $n=50$, $n=100$ and $n=150$. The time has been linearly rescaled so that the root is at $t=1$ and the current time is 0. (B) The upper (for the neutral curve) and lower (for the non-neutral curves) $\%95$ confidence intervals for the curves in part (A). In all cases, $N=64000$, $\epsilon=0.1$ } \label{test lin} \end{figure*}

To each leaf $i$ in a tree, a number $N_i$ is assigned. This is the number of internal nodes between leaf $i$ and the root. The  variance of this number in a tree is given by  $\sigma^2_n=\frac{1}{n}\sum_{i=1}^{n}(N_i-\bar{N})^2$. In a completely symmetric tree $\sigma^2_n=0$. Fig. \ref{test sigma}A presents the distribution of $\sigma^2_n$ for both neutral and non-neutral trees for three sample sizes. As expected, the results indicates that the coalescent trees in the presence of selection are, on the average, more asymmetric compared to neutral trees. In addition, as the sample size increases, the distribution of $\sigma^2_n$ differs more between the neutral and non-neutral cases. However, even for sample size $n=200$, there is a significant overlap between the neutral and non-neutral distributions. Therefore, this measure is not a useful test to detect a tree signiÞcantly distinct from the neutral expectation. Similar conclusion was reached in \cite{maia2004effect} where authors have analyzed three more measures than the one presented here.

The above measure does not take into account the information about the branch-length in a tree. We have looked at two quantitates which use this information. Fig. \ref{test sing}A shows the expected number of singleton lineages left in a tree as a function of time for three sample sizes. These curves are obtained by averaging over random samples and over population replicas. The curve for the neutral case falls clearly below the rest of the curves. To see if the separation between the neutral and non-neutral curves is large enough that one can differentiate whether or not a single tree is neutral, we also show the confidence intervals in Fig. \ref{test sing}B. The upper $95\%$ confidence interval for the neutral case falls below the lower $\%95$ confidence for almost all the cases with selection. The separation becomes larger as the sample size increases. Even for the parameter combination where the dynamics falls at the boundary between the multisite selection and the selective sweep regime ($N=64000$ , $\mu=10^{-4}$, $s=10^{-3}$ and $\epsilon=0.1$), the lower confidence interval is very close to the upper confidence interval for the neutral curve. Fig. \ref{test lin}A shows the average number of lineages left in a tree as a function of time. The confidence intervals are also shown in Fig. \ref{test lin}B. Again, for $n=100$ or $n=200$, the upper confidence interval curve for the neutral case falls below the lower confidence interval curves for all the cases with selection. 

\subsection{Correlation between weight and fitness of ancestors}
In Fig. 4A of the main text, we showed the distribution of the fitness of the ancestors for certain time intervals in the past for a set of parameters. Let us denote this distribution by $\phi_t(f)$. In the limit of large times, this distribution is equal to the fitness distribution for the common ancestor of the whole population, $\phi_{\infty}(f)$. In Fig. 4B, we also showed the scatter plot between the weight of ancestors and their fitness advantage for $t=100$ generations in the past. The scatter plot represents the joint distribution of weight and fitness of ancestors, $\phi_t(f,w)$. 

\begin{figure*}[btp]
\includegraphics{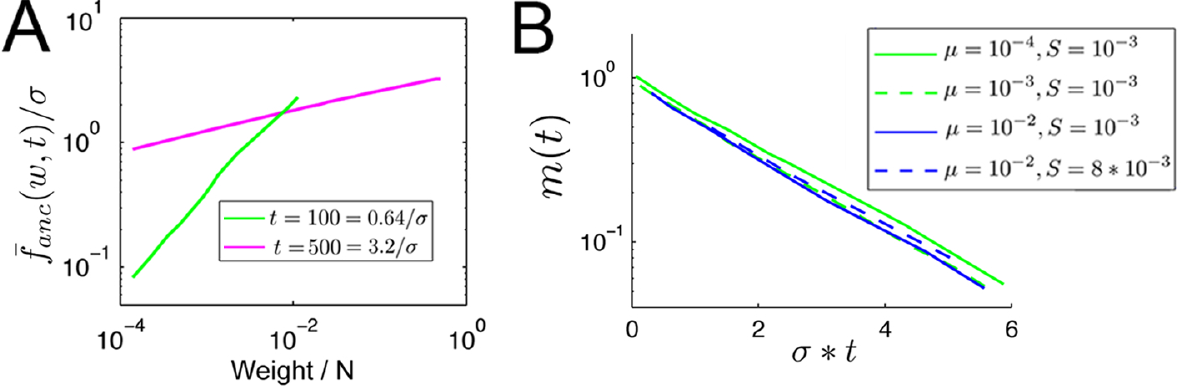}
 \caption{Correlation between the weight and fitness of ancestors. (A) Average fitness of an ancestor conditional on its weight for two time intervals. Note the log-log scale. $N=64000$, $\epsilon=0.1$, $\mu=10^{-3}$ and $s=8*10^{-3}$. (B) Fitting a line to the curve in part (A) gives a time dependent slope $m(t)=\log(\bar{f}_{anc}(w,t))/\log(w)$. The slope $m(t)$ is plotted as a function of time for a few different parameter values. Note the log scale on the y-axis. The time for each parameter set has been rescaled by the corresponding $\sigma$. Sample size $n=100$, $N=64000$ and $\epsilon=0.1$. \label{fitness_dis_time_supl} } \end{figure*}

 One can consider the expected fitness of an ancestor given its weight, $\bar{f}_{anc}(w,t)=\sum\limits_{f} f * \phi_t(f|w)$. Fig. \ref{fitness_dis_time_supl}A shows $\bar{f}_{anc}(w,t)/\sigma$ as a function of $w/N$ for $t=100$ and $t=500$ in log-log scale.  The dependence seems to be linear, namely, $\bar{f}_{anc}(w,t)\propto w^{m(t)}$, where $m(t)$ is the slope of the lines in Fig. \ref{fitness_dis_time_supl}A. This slope depends on the time, and of course, other parameters such as $N, \mu$, etc. Fig. \ref{fitness_dis_time_supl}B shows $m(t)$ as a function of time for different sets of parameters. For each set of parameters, the time axis has been rescaled with the fitness variance for the corresponding parameter set, $\sigma(N, \mu,\epsilon,s)$.   As we see, the slope $m(t)$ drops as a function of time. In other words, the correlation between the weight and the fitness of ancestors reduces as one goes further back in time.

 \begin{figure*}[btp]
\includegraphics{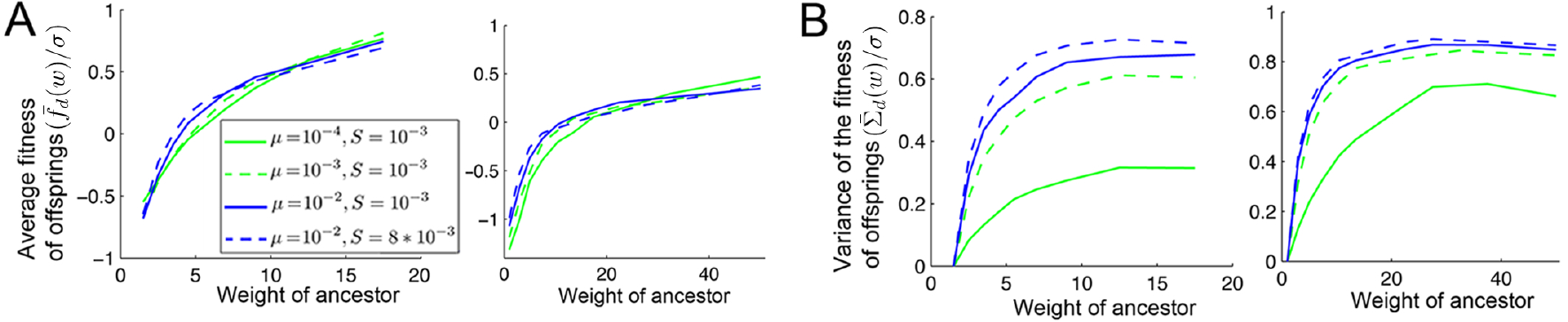}
 \caption{Correlation between the weight and fitness of offspring. (A) Average fitness of genomes as a function of the ancestral weight for two different time slices in the past. (B) Variance in the fitness of genomes as a function of the ancestral weight for two different time slices in the past.  $N=64000$, $\epsilon=0.1$ in all cases. \label{offspring_fitness_supl} } \end{figure*}
 
In the main text, we also presented some results on the relation between the weight of an ancestor in a tree, $w_i$, and the fitness of the $w_i$'s genomes in the sample which are derived from that ancestor.  In particular, we focused on the mean,  $f_d(w_i)=\frac{1}{w_i}\sum\limits_{j=1}^{w_i} f_{j}$,  and the variance, $\Sigma_d^2(w_i)={\frac{1}{w_i}\sum\limits_{j=1}^{w_i}( f_{j}-f_d(w_i))^2}$ (see below for an example of a tree explaining the notation). The average of these quantities over random samples of genomes and over population replicas are denoted by $\bar{f}_d(w_i)=<\frac{1}{w_i}\sum\limits_{j=1}^{w_i} f_{j}>$ and $\bar{\Sigma}_d^2(w_i)=<\frac{1}{w_i}\sum\limits_{j=1}^{w_i}( f_{j}-f_d(w_i))^2>$.

In the main text, we only presented these quantities for two parameter sets. In Fig. \ref{offspring_fitness_supl}A and B, we show $\bar{f}_d(w)/\sigma$ and $\bar{\Sigma}_d^2(w)/\sigma$ for more parameter sets. The sample size is $n=100$ and the results are shown for two different time points. One of the time points is chosen to be the first time that the tree carries a lineage with weight greater than $15\%$ of the sample size. The other time point corresponds to the first time the weight of a single lineage becomes greater than $40\%$ of the sample size. 
 
\subsection{Fitness proxy score and its performance}
Consider a sample of $n$ genomes and the corresponding reconstructed phylogenetic tree. Although there is always a positive correlation between the weight of an ancestor and its fitness and the fitness of its derived genomes, both of these correlations drop as one goes further back in time. When most of the lineages have condensed into high-weight ancestors, the average fitness of the offspring of such ancestors is close to zero and there is little correlation between the weight and the fitness of the offspring (see right plot in Fig. 4C of the main text). The variance in the fitness of the offspring also becomes close to the population fitness variance $\sigma$. In other terms, all of the derived genomes of such high-weight ancestors are, more or less, evenly distributed across the fitness distribution. 

This is consistent with our observations in Fig. 4D of the main text. As the coalescent time for a pair of genomes increase, the difference in the fitness of the two genomes increases as well. This means,  as $\tau_{ij}$ becomes larger compared to $T_2$ (the region covered in yellow and red colors in Fig. 4D), there is less information about the fitness of the pair of genomes involved. For example, one can have high fitness and the other one low fitness, or both can have average fitness. In other words, when the coalescent time for a pair of genomes becomes larger compared to the population average $T_2$, there is more uncertainty on the fitness of that pair of genomes. 

Because of the above argument, we do not want the scoring scheme to be affected by the coalescent events far back in the tree. In addition, as we saw in Fig. 4D of the main text, when the fitness of two genomes is higher, the coalescent time between them is shorter compared to the mean pairwise coalescent time for the whole population $T_2$. The correlation between weights and fitness is also stronger for earlier times. Therefore, the earlier a coalescent event, the more it should affect the scores. It is important to have a sense of `early times' or `late times' in a tree. We use the empirical value of the mean pairwise coalescent time (i.e. estimate of $T_2$ from the sample) for this purpose. In the algorithm, the time values appear only in the form of ratios. So having a correct estimate of the mutation rate is irrelevant.

\begin{figure}[btp]
\includegraphics{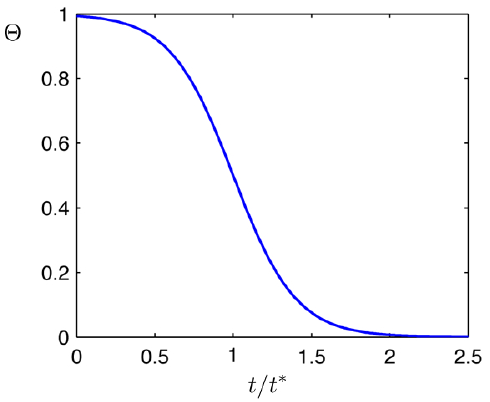}
 \caption{The Fermi-Dirac function $\Theta(t)=\big(1+\exp(5 \times (t/t^*-1))\big)^{-1}$. \label{FD}} \end{figure}

\begin{figure}[btp]
\includegraphics{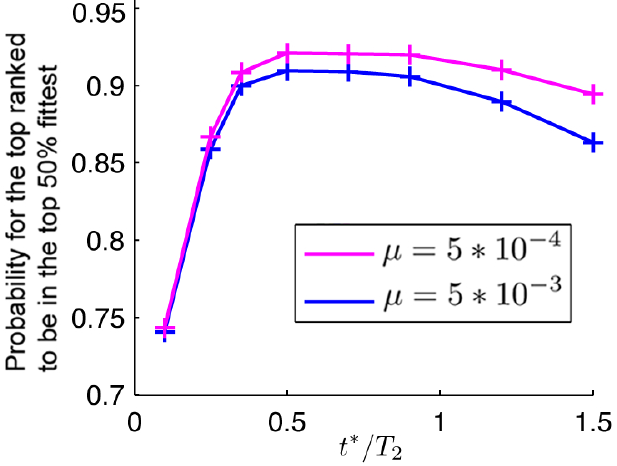}
 \caption{Performance as a function of the threshold $t^*$. $N=64000$, $\epsilon=0.1$ and $s=2*10^{-3}$. \label{percent_mean_teffect}} \end{figure}

To incorporate the above ideas, we introduced a threshold time $t^*=x_* \times T_2$ and have the coalescent events which happen at a time further back compared to $t^*$ contribute progressively less on the score. On the other hand, the coalescent events earlier than this stage will be progressively more important in the scoring scheme. In order to do this, we introduced the function $\Theta(t)$ with a Fermi-Dirac form, shown in Fig. \ref{FD}. In the results shown in this paper on the performance of the algorithm, we set $t^*=0.5*T_2$, where $T_2$ is the average pairwise coalescent time. We checked the performance for various values of $t^*$. We found that, in general, the results are very robust within a range of $0.4*T_2 < t^* < T_2$. Outside this range the performance slightly decreases. For the sake of example, in Fig. \ref{percent_mean_teffect}, we show the probability for the fitness of a genome within the top \%10 ranked to belong to the top $50\%$ fitness values as a function of the threshold parameter, $t^*$, for two different mutation rates. 

 
\begin{figure*}[btp]
\includegraphics{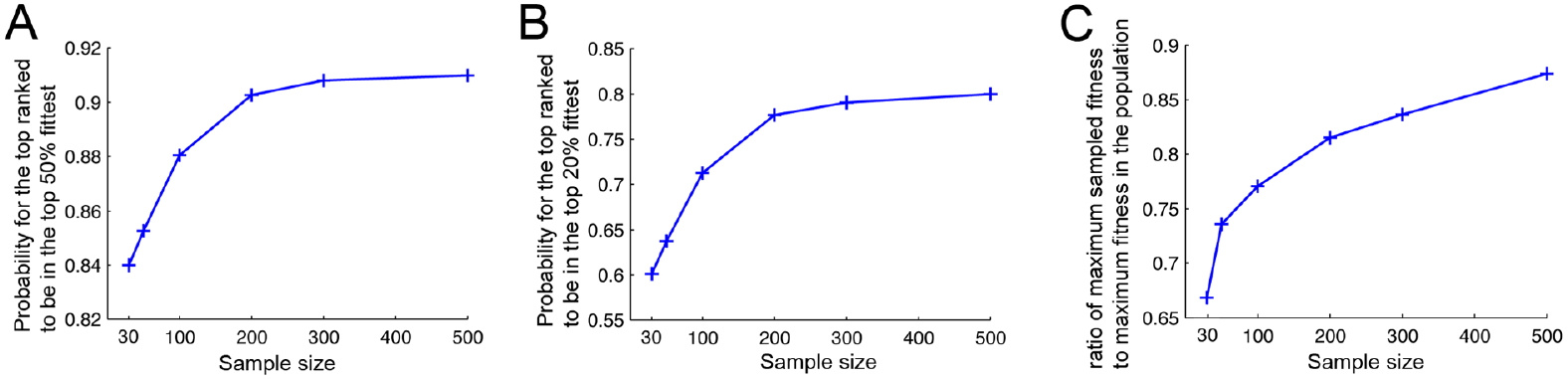}
 \caption{Performance as a function of the sample size. $N=64000$, $\epsilon=0.1$, $\mu=5*10^{-3}$ and $s=2*10^{-3}$. \label{sizeeffect}} \end{figure*}

We have also evaluated the performance of the algorithm for different sample sizes. In Fig. \ref{sizeeffect}, we present the results for a set of parameters.  We see that for samples smaller than $n=100$. the performance decreases, whereas for higher samples sizes, the performance is similar to the results shown above for $n=200$. For example, for a sample of size $n=30$, and for parameters $\mu=5*10^{-3}$ and $s=2*10^{-3}$, the probability for the fitness of the top ranked genome to belong to the top $50\%$ values turns out to be around 0.84, compared to 0.9 for sample size of $n=200$. 

Another point is that, as sample size becomes smaller, the right tail of the fitness distribution (see Fig. 2A and B) becomes under sampled. It has been shown that in similar models as the one we have considered here, the bulk of the fitness distribution can be approximated by a Gaussian profile \citep{desai2007beneficial}. For a Gaussian distribution, the probability of sampling a point with value of at least one (two) $\sigma$ above the mean is around 0.15 (0.3). By inspecting the fitness profiles in Fig. 2A and B of the main text, as well as profiles shown in Fig. \ref{more profiles} of SI, we see that the frequency of clones with fitness more than one $\sigma$ above the population average is around $0.1$.  This frequency for clones with fitness more than $2\sigma$ above the population average is less than $p=0.05$. In Fig. \ref{sizeeffect}C, we see the ratio of the maximum fitness value in a sample of size $n$ to the maximum fitness value that exist in the population. As expected, the larger the sample size, this ratio gets closer to one.

\begin{figure*}[btp]
\includegraphics{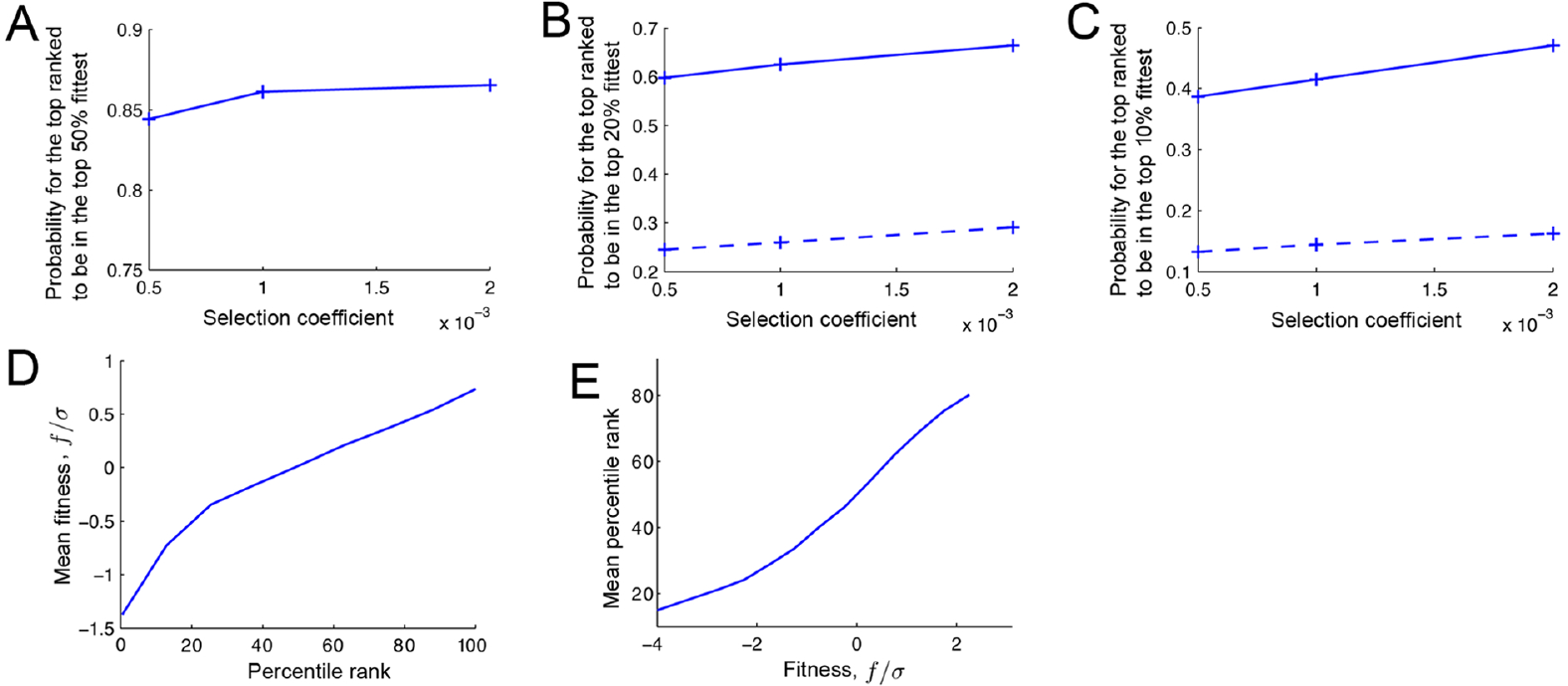}
 \caption{Performance of the fitness ranking algorithm in the case of purifying selection. Sample size $n=200$, $N=32000$ and $\mu=5*10^{-3}$ in all plots. (A) Probability for the fitness of a genome within the top 10\% ranked to belong to the top $50\%$ fitness values. (B) Probability for the fitness of a genome within the top 10\% ranked to belong to the top $20\%$ fitness values. The dashed line shows this probability for a randomly chosen genome. (C) Probability for the fitness of a genome within the top 10\% ranked to belong to the top $10\%$ fitness values. (D) Mean fitness as a function of the rank. Selection coefficient $s=10^{-3}$. (E) Mean rank as a function of the fitness. Selection coefficient $s=10^{-3}$. \label{purifying}}\end{figure*}
 
\begin{figure*}[btp]
\includegraphics{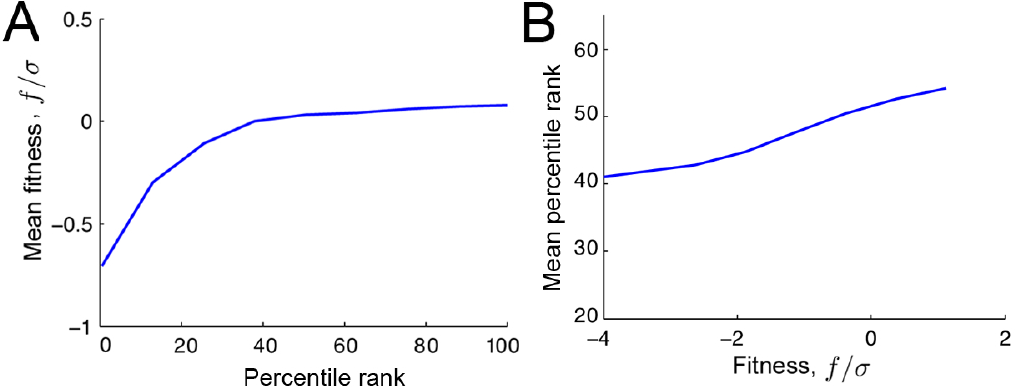}
 \caption{Performance of the fitness ranking algorithm in the case of purifying selection decreases for small values of $\mu_d/s$. Sample size $n=200$,  $s=2*10^{-3}$, $\mu=10^{-3}$, $N=32000$ and $\epsilon =  0$ in both plots. (A) Mean fitness as a function of the rank.  (B) Mean rank as a function of the fitness. \label{purifying failure}}\end{figure*}
 
\begin{figure*}[btp]
\includegraphics{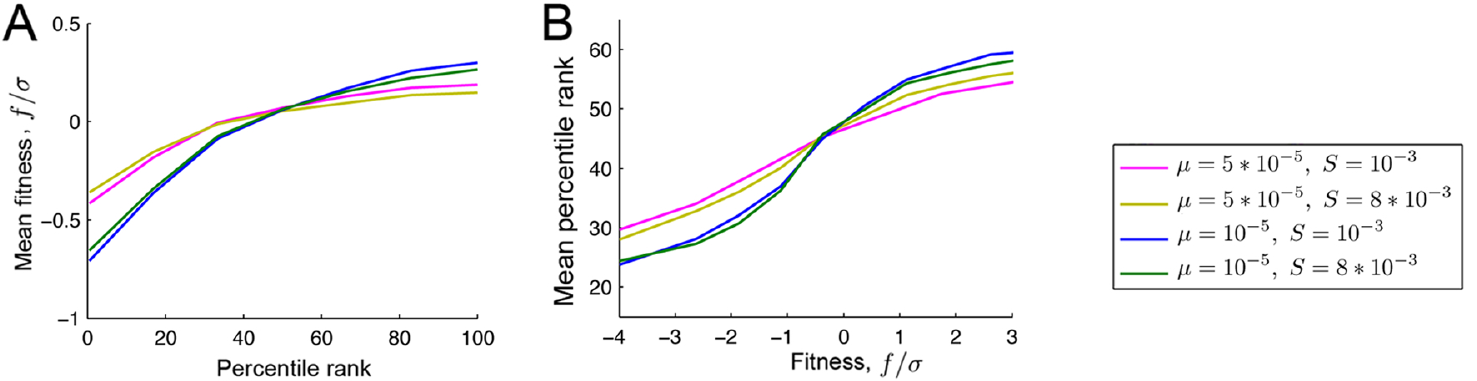}
 \caption{Performance of the fitness ranking algorithm decreases for small values of the populations size and the beneficial mutation rate, $\mu_b=\epsilon*\mu$, as the system approaches the regime of selective sweeps/successive mutations. Sample size $n=200$, $N=16000$ and $\epsilon =  .1$ in both plots. (A) Mean fitness as a function of  rank.  (B) Mean rank as a function of  fitness. \label{failure}}\end{figure*} 

 \begin{figure}[btp]
\includegraphics{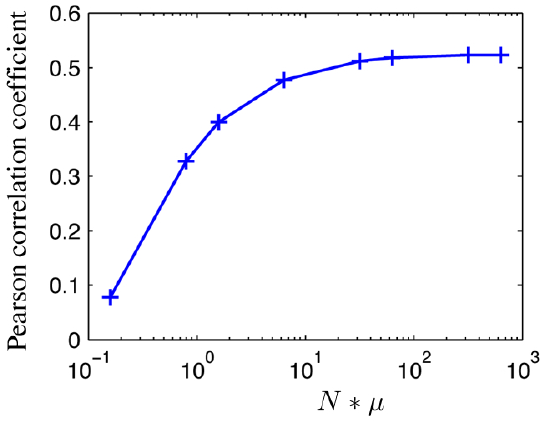}
 \caption{Correlation between the rank and the distance $d^\prime$ (Hamming distance to the ancestors of the future generations in the current population)  as a function of $N*\mu$. For small population sizes and mutation rates where $N*\mu<1$, the correlation becomes smaller and the performance of the fitness ranking algorithm decreases. $\epsilon =  0.1$ and $s =  0.002$.  \label{pearson}}\end{figure} 
  
We have  studied the performance of the algorithm in the presence of purifying selection. i.e. when $\epsilon =  0$. The results presented in Fig. \ref{purifying} show that the algorithm performs well in this regime, similar to the case of adaptation (i.e. $\epsilon =  0.1$). 

As we have mentioned in the main text, we are interested in the sets of evolutionary parameters for which several mutations segregate simultaneously and the population is formed by several clones with varying fitness values. In the opposing limit  corresponding to the regime of selective sweeps/successive mutations, we expect the performance of the algorithm to deteriorate, as some of the fundamental aspects of the dynamics (such as the dependence of the fate of mutations on the genetic background) are different. Below we show results on how the performance of the algorithm decreases for parameter combinations which do not satisfy  the above condition.

In the case of purifying selection where deleterious mutations are present ($\epsilon=0$), the validity of our assumption depends on the inequality $N\exp(-\mu_d/s)<<\frac{1}{s}\log(\mu_d/s)$, where $\mu_d$ is the deleterious mutation rate and $s$ is the deleterious effect of mutations. By comparing the results presented in Fig. \ref{purifying failure}, with the ones shown in Fig. \ref{purifying}D and E, we see how the algorithm performs inferior for the parameter combination not satisfying the above condition. 

In the presence of beneficial mutations, the condition for the validity of our assumption is given by the inequality $N\mu_b s  >>  s/(1+\log(Ns))$. By comparing the results presented in Fig. \ref{purifying}, with the ones shown in Fig. 6A and B of the main text, we see how the performance of the algorithm decreasease for some parameter combinations that do not satisfy the above condition. To make this point clear, we calculated the Pearson correlation coefficient between the rank and the distance $d^\prime$ -- as in the main text, $d_i^\prime$ is the the average of Hamming distance from individual $i$ in the sample to all of the genomes in the current population that are direct ancestors of the population in the future (we know these ancestors from the forward simulation). In Fig. \ref{purifying}, we show this correlation as a function of the parameter $N*\mu$. As we see, for smaller values of this compound parameter, particularly for $N*\mu<1$, the correlation coefficient drops.

\newpage 
.
\newpage 
.

 
\end{document}